\documentclass[12pt,preprint2]{aastex}
 \def\ltsima{$\; \buildrel < \over \sim
\;$} \def\simlt{\lower.5ex\hbox{\ltsima}}            
\def\gtsima{$\; \buildrel > \over \sim \;$}
\def\simgt{\lower.5ex\hbox{\gtsima}}            

\def\oviii{{\sc O\ viii}}
\def\ovii{{\sc O\ vii}}

\def\civ{{\sc Civ\/}}
\def\nv{{\sc Nv\/}}
\def\ovi{{\sc Ovi\/}}

\begin{document} 

\title{\sc Toward a self consistent model of the ionized absorber in NGC 3783}
\author{Y. Krongold\altaffilmark{1}, F. Nicastro\altaffilmark{1}, N.S. Brickhouse\altaffilmark{1},
M. Elvis\altaffilmark{1} , D.A. Liedahl\altaffilmark{2}, and S.
Mathur\altaffilmark{3} }

\altaffiltext{1}{Harvard-Smithsonian Center for Astrophysics. 60
Garden Street, Cambridge MA 02138}

\altaffiltext{2}{Lawrence Livermore National Laboratory,
Department of Physics and Advanced Technologies, 7000 East Avenue,
L-473, Livermore, CA 94550} \altaffiltext{3}{Department of
Astronomy, Ohio State University, 140 West 18th Avenue, Columbus,
OH 43210}


\begin{abstract}

We present a detailed model for the ionized absorbing gas evident
in the 900 ksec {\it Chandra} HETGS spectrum of NGC 3783. The
analysis was carried out with PHASE, a new tool designed to model
X-ray and UV absorption features in ionized plasmas. The 0.5-10
keV intrinsic continuum of the source is well represented by a
single power law ($\Gamma=1.53$) and a soft black-body component
(kT $\sim\ 0.1$ keV). The spectrum contains over 100 features,
which are well fit by PHASE with just six free parameters. The
model consists of a simple two phase absorber with difference of
$\approx 35$ in the ionization parameter and difference of
$\approx 4$ in the column density of the phases. The two
absorption components turned out to be in pressure equilibrium,
and are consistent with a single outflow ($\approx 750$ km
s$^{-1}$) and a single turbulent velocity (300 km s$^{-1}$), and
with solar elemental abundances. The main features of the low
ionization phase are an Fe M-shell unresolved transition array
(UTA) and the \ovii \ lines. The \ovii \ features, usually
identified with the \oviii \ and a warm absorber, are instead
produced in a cooler medium also producing O {\sc vi} lines. The
UTA sets tight constraints on the ionization degree of the
absorbers, making the model more reliable. The high ionization
phase is required by the \oviii \ and the Fe L-shell lines, and there
is evidence for an even more ionized component in the spectrum. A
continuous range of ionization parameters is disfavored by the
fits, particularly to the UTA. Our model indicates a severe
blending of the absorption and emission lines, as well as strong
saturation of the most intense O absorption lines. This is in
agreement with the \ovii \ ($\tau_{\lambda}=0.33$) and \oviii \
($\tau_{\lambda}=0.13$) absorption edges required to fit the
spectrum. The low ionization phase can be decomposed into three
subcomponents, based on the outflow velocity, FWHM, and H column
densities found for three out of the four UV absorbers detected in
NGC 3783. However, the ionization parameters are systematically
smaller in our model than derived from UV data, indicating a lower
degree of ionization. Finally, our model predicts a Ca {\sc xvi}
line for the feature observed at around 21.6 \AA \ (a feature
formerly identified as \ovii), constraining the contribution from
a zero redshift absorber.

\end{abstract}

\keywords{galaxies: absorption lines -- galaxies: Seyferts --
galaxies: active -- galaxies: X-ray}
\section{Introduction}

Quasars and Active Galactic Nuclei (AGNs) display a rich and
confusing array of atomic emission and absorption features
throughout the infrared, optical, ultraviolet and soft X-ray
bands. Does this profusion of features reflect an inherent
complexity or even randomness within AGNs? Or is there an
underlying order that is masked by the overabundance of detail?
The case has been made on both sides (for complexity: Baldwin et
al. 1996; Korista et al. 1996; Krolik \& Kriss 2001; for
underlying order: Mathur, Elvis \& Wilkes 1995; Elvis 2000; Arav
et al. 2003).

One case has been intensively pursued lately, due to the
improvement in the quality of UV and X-ray spectra from STIS,
FUSE, and the grating spectrometers on {\em Chandra} (HETGS,
LETGS) and XMM-Newton (RGS): the highly ionized (or `warm')
absorbers that appear in about half of all UV and X-ray spectra of
AGNs.  Both UV and X-ray absorption lines show moderate velocity
($\sim$1000~km~s$^{-1}$) blueshifts, implying outflows.  The mass
loss rate from these outflows can be a substantial fraction of the
accretion rate needed to power the AGN continuum, and so these
`quasar winds' are likely to be dynamically important to AGNs
(Mathur, Elvis \& Wilkes 1995). Understanding the physical state
or states of this gas, and hence the location and dynamics of AGN
winds would be a valuable advance in the understanding of quasars.

The UV and X-ray absorbers must be closely related: exactly the
same 50\% of AGNs show both \citep{ma95,ma98,cren99,mo01}, and
each can be used to predict the presence of the other.  However,
the absorbers show complex, multi-component and time-variable
structures in UV spectra, and it is not straightforward to connect
these to the X-ray lines, given the order of magnitude lower
spectral resolution in the X-ray band. Moreover column densities
derived from the UV lines are strongly affected by scattered light
into our beam or by partial covering of the continuum source (Arav
et al. 2003). While the UV absorbers are dominated by O {\sc vi} lines,
with C {\sc iv} and N {\sc v} also appearing, the X-ray absorbers
are dominated
by \ovii \ and \oviii, with many other lines of similar ionization now
being seen. Despite being dubbed `warm absorbers' the absorbers
respond to changes in the ionizing continuum and so are at least
in part photoionized \citep{nic99}. Hence we shall use the less
loaded term `ionized absorbers'.

Here we take the best existing high resolution X-ray spectrum of
an ionized absorber in an AGN, a 900~ksec observation of NGC~3783
with the {\em Chandra} HETGS, and analyze the absorption features
self-consistently with a new code (`PHASE', Krongold et al. 2003,
in preparation). We find a strikingly simple solution: well over
100 absorption features can be fitted by a two component solution
having only six free parameters. Moreover, allowing for the lower
X-ray resolution, the lower ionization component is consistent
with two UV absorbers. There is also evidence in the X-rays for a
third, high velocity and low column density, UV component in four
lines. We discuss in detail why we are able to obtain this simple
solution, in contrast to earlier studies of the same object and
spectrum that led to more complex answers.  The two component
solution is strongly suggestive of a two-phase medium in pressure
balance, with further likely implications.

\subsection{High Resolution X-ray Spectra of NGC~3783}

NGC 3783 is an extensively studied bright Seyfert galaxy, at a
redshift 0.00976 $\pm 0.00009$ (2926 $\pm 28$ km s$^{-1}$, de
Vaucouleurs et al. 1991).
NGC 3783 has been the subject of extensive monitoring by the  High
Energy Transmission Grating Spectrometer (HETGS, Canizares et al.
2000) on board the {\it Chandra X-ray Observatory}, with a total
exposure of $\approx$ 900 ksec (reported in Kaspi et al. 2002).
The resulting spectrum covers the 0.5-10 keV energy range, and is
the best obtained so far for an ionized absorber in a Seyfert
galaxy, with more than 2000 counts per resolution element at 7
\AA. This spectrum showed the presence of more than one hundred
absorption lines from a wide range of ionized species. A few weak
emission lines, mainly from O and N, were also detected. The
spectrum further revealed  a wide absorption feature in the
$16-17$ \AA \ range. \citet{ka02} identified this feature as an
unresolved transition array (UTA), arising from numerous Fe
M-shell inner-shell $2p-3d$ transitions. The XMM-Newton RGS has
also observed NGC 3783 (Blustin et al. 2002). This observation
extends to longer wavelengths than the 900 ksec exposure from the
HETGS. A complete review of the analyses carried out on these data
will be presented in \S \ref{comp}. To this date, self consistent
modelling for the full 900 ksec {\em Chandra} high resolution
spectrum has not been published.

NGC 3783 has also been studied in the UV band with FUSE and
HST-STIS. These observations have revealed intrinsic absorption by
Ly$\alpha$, O {\sc vi}, and N {\sc v} in this region of the
spectrum (Kraemer, Crenshaw \& Gabel, 2001, and references
therein). The high spectral resolution and signal to noise ratio
of both data sets allowed the identification of four absorption
systems moving toward us, with outflow velocities of 1320, 1027,
724, and 548 km s$^{-1}$ (Gabel et al. 2003).

In this paper, we present the first self-consistent model for the
900 ksec X-ray absorption spectrum of NGC 3783. In the following
section (\S \ref{dat}) we describe the data reduction. The
modelling was carried out with ``PHASE'', a newly developed code
based on CLOUDY (v. 90.04, Ferland 1997) and the Astrophysical
Plasma Emission Database (APED v. 1.3.0, Smith et al. 2001), and
designed to reproduce absorption features produced by ionized
plasmas. This code is briefly explained in section \S \ref{code}.
In \S \ref{fit} we present the results obtained for the spectral
fitting and in \S \ref{disc} and \S \ref{disc2} we discuss the
reliability of the model and its implications.

\section{Data Reduction \label{dat}}

\begin{figure}[!t]
\figurenum{1} \plotone{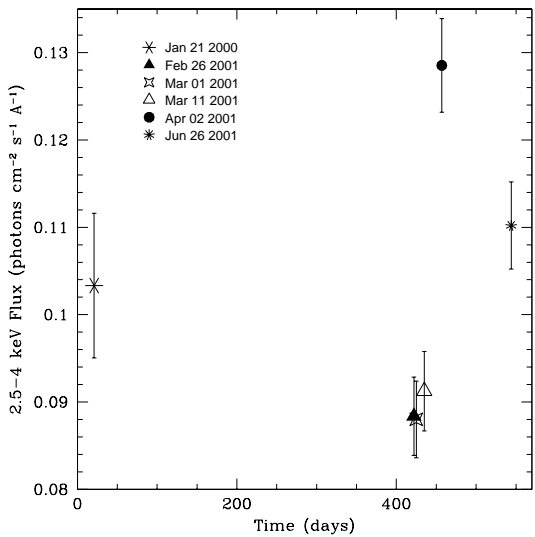} \caption[f1.eps]{Light curve of
Chandra-HETG observation of NGC3783 in the 2.5-4 keV (3-5 \AA)
range. \label{var}}
\end{figure}

NGC 3783 was observed six times using the HETGS on board the
Chandra X-ray Observatory, with the Advanced CCD Imaging
Spectrometer (ACIS). The first observation was carried out on
January 2000 and had a duration of 56 ksec (reported in Kaspi et
al. 2000). Between February and June of 2001, the following five
observations were taken, each one with an exposure time of 170
ksec (reported in Kaspi et al. 2002). Figure \ref{var} presents
the light curve in the 2.5-4 keV ($\sim$ 3.0-5.0 \AA) range. We
chose this range because it avoids the Fe K 6.4 keV complex and
most absorption from bound-free transitions. Therefore, it can
indicate changes from the central source. In agreement with
\citet{ka02}, we found variability at a level of about 50\%
between the fourth and fifth observations, and smaller in the rest
of the cases. In this paper we present a model for the averaged
spectrum of the six observations. We do not expect the level of
variability to affect our main results. However, a detailed
analysis on each observation, and the general effects of
variability will be presented in the future.

We retrieved from the public Chandra data archive\footnotemark
\footnotetext{http://asc.harvard.edu/cda} the primary and
secondary data products for these six observations and reprocessed
their event files with the Chandra Interactive Analysis of
Observations (CIAO\footnotemark
\footnotetext{http://asc.harvard.edu/ciao}) software (Version
2.3). Following the on-line data analysis ``threads'', provided by
the Chandra X-ray Center (CXC\footnotemark
\footnotetext{http://asc.harvard.edu/ciao/documents\_threads.html}),
we extracted from the CALDB\footnotemark \ \citep{gra96}
\footnotetext{http://cxc.harvard.edu/caldb} the source spectra and
the corresponding 1st-order Ancillary Response Files (ARFs) and
Redistribution Matrices (RMFs).

\begin{figure}[!t]
\figurenum{2} \plotone{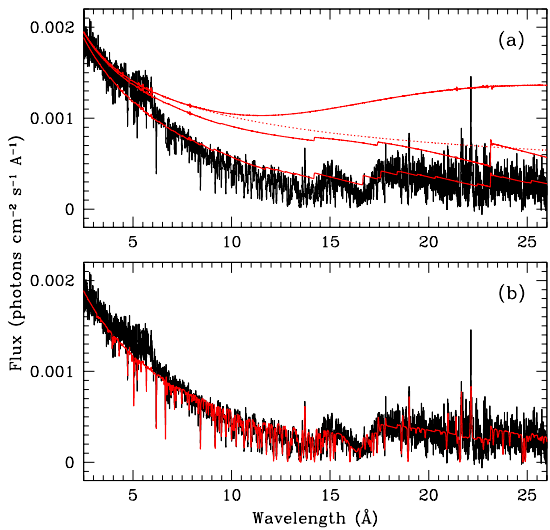} \caption[f2.eps]{Fluxed MEG
first-order spectrum of NGC 3783. (a) Continuum fitted to the
spectrum. The three solid lines stand for (from upper to lower):
intrinsic continuum of the source, continuum attenuated by
Galactic absorption, and continuum further attenuated by the two
phase absorber edges. The dotted line represents the predicted
power law without the contribution from the black body component.
(b) Two phase absorption model plotted for comparison.$^9$

$^9${\footnotesize Figures  available in color in the electronic
format.} \label{flux}}
\end{figure}

The HETGS (Canizares et al. 2000) produces spectra from two
grating assemblies, the high-energy (HEG) and medium energy (MEG)
gratings. All the spectra described below and throughout the paper
correspond to the MEG. We combined the -1st and +1st orders for
each exposure. To further increase the signal to noise ratio of
the data, we coadded the spectra and averaged the ARFs, weighting
them according to their respective exposure times. The resulting
spectrum has a net exposure of 888.7 ksec. The signal to noise
ratio per resolution element ($\approx$0.02 \AA \ for the MEG)
varies from $\approx 46$ to $\approx 5$ in the 6-25\AA \
($\approx$ 0.5-2.0 keV) range. The fluxed spectrum is presented in
Figure \ref{flux}a, and the empirical spectrum is presented in
Figure \ref{fig1}.

\section{Modelling: PHASE \label{code}}

With the aim of modelling in a self-consistent way all the
absorption features observed in the spectra of UV and X-ray
sources, we have developed PHASE (PHotoionized Absorption Spectral
Engine), a code that calculates absorption due to an ionized
plasma based on ATOMDB\footnotemark \ (v.1.3.0, Smith et al.
2001\footnotetext{http://cxc.harvard.edu/atomdb}) and CLOUDY (v.
90.04, Ferland 1997). The calculations assume a simple geometry
that consists of a central source emitting an ionizing continuum
with clouds of gas intercepting our line of sight. The ionization
balance is calculated using CLOUDY with an SED chosen to be
consistent with the intrinsic Spectral Energy Distribution (SED)
of the source. The parameters of the code are (1) the ionization
parameter (defined over the entire Lyman continuum, as
$U=Q(H)/4\pi r^2 n(H) c$ with $Q(H)$ the number of H ionizing
photons per second, $r$ the distance to the source, $n(H)$ the H
density, and $c$ the speed of light.), (2) the equivalent hydrogen
column density, (3) the outflow velocity, and (4) the internal
micro-turbulent velocity of each one of the absorbers, as well as
(5) the intrinsic SED of the source. A sixth parameter, the
electron temperature, can be included, forcing the temperature to
be different from photoionization equilibrium. PHASE can thus
handle hybrid states, e.g. gas photoionized by a strong radiative
source but where another external mechanism (for instance, shock
heating) keep the electron temperature at a higher value than the
one expected only through photoionization equilibrium. The
micro-turbulent velocity and the thermal velocity are added in
quadrature to compute the total Doppler broadening parameter of
the gas: $V_{DOP}^2=V_{Therm}^2+V_{Turb}^2$.

The code proceeds as follows: (1) The absorption produced by each
transition of the different ion species is calculated, using the
formula
$$\tau_{\nu}\equiv{\int_{0}^{L}}ds\alpha_{\nu}=N_{ion}\frac{\pi
e^2}{mc}f_{lu}\Phi_\nu$$ where $\alpha_{\nu}$ is the absorption
coefficient at the frequency $\nu$, $L$ is the linear size of the
cloud, $N_{ion}$ is the column density of the given ion
(calculated through the ionization balance), $m$ is the electron
mass, $e$ is the electron charge, $c$ is the speed of light,
$f_{lu}$ is the oscillator strength of the electron transition
(from the lower to the upper level), and $\Phi_\nu$ is the line
profile (which is assumed to be a Voigt profile, Rybicki \&
Lightman, 1979, and calculated by convolving the natural
broadening with the thermal and microturbulent Doppler
broadening).

(2) Then the model calculates the continuum absorption produced by
bound-free transitions.

(3) The contributions from different absorption lines and edges
are added together to produce an absorption spectrum.

(4) This spectrum is finally shifted to the proper outflow
velocity.

PHASE includes $\sim4000$ lines arising from the ground level. The
line list was extracted from the Astrophysical Plasma Emission
Database (APED v. 1.3.0, Smith et al. 2001). APED includes
information up to the 5th principal quantum number for transitions
arising from the 14 more abundant elements in the universe, namely
H, He, C, N, O, Ne, Mg, Al, Si, S, Ar, Ca, Fe, and Ni (for a
detailed list of the ions of each element included in APED visit
the ATOMDB web page, http://cxc.harvard.edu/atomdb). To further
increase the number of lines
in our model, we added nearly 200 features involving principal
quantum numbers n=5 to n=10 (levels not present in APED) from the
``Atomic Data for Permitted Resonance Lines'' \citep{ver96} . This
compilation includes lines connecting the ground level to excited
levels via optically allowed transitions, i.e. transitions between
levels of opposite parity with $|\Delta L| \leq 1$, $|\Delta
S|=0$, and $|\Delta J| \leq 1$, where $L$ is the orbital, $S$ the
spin, and $J$ the total angular momentum quantum number. In order
to include in our calculations the Fe M-shell inner transitions
(which are responsible for the UTA), we used the abbreviated data
set provided by \citet{behar01}, the best source of information
for these transitions available in the literature. This
abbreviated version of the data assumes that mean wavelengths and
effective oscillator strengths can emulate the complex absorption
produced by several lines of each ion. According to the authors,
the approximation can successfully reproduce the general shape and
EW of the feature. The only exception is for Fe {\sc x-xiii} ions;
in this case, if the ion column density lies in the range
$10^{17}-10^{18}$ cm$^{-2}$ and the turbulent velocity is
$\lesssim100$ km s$^{-1}$, the approximation can overestimate the
actual value of the EW by as much as 40\%. We also included the
most intense (oscillator strengths $> 0.1$ m\AA) inner shell
(1$s$-2$p$) transitions from Ne, Mg, Al, Si, S, Ar, Ca, and Fe.
These data were extracted from the list of \citet{behar02}, and
will be included in APED (v.2.0.0). A full explanation of the
present code will be published in a forthcoming paper (Krongold et
al., in preparation).

We performed spectral fitting of the NGC 3783 data using PHASE
integrated into the Sherpa (Freeman et al. 2001) package in CIAO
(Fruscione2002) as a table model.

\section{Spectral Fitting \label{fit}}

The key point of our modelling was to fit self-consistently the
absorption lines and edges from all the ions of all the species at
the same time, without any assumption about the absorption of
individual ions. Therefore, we did not introduce any prior
constraint on the column density or population fraction of any
ion. As will be discussed in \S \ref{disc}, many lines of
different ions are blended, making the measurement of individual
ion column densities unreliable in most cases. The determination
of individual ion column densities through continuum bound-free
absorption is unreliable also due to the contribution of so many
different ions. Furthermore, even if the column densities produce
edges, these might be hidden by absorption lines. Such a case is
clearly exemplified in Figure \ref{smita} for the spectrum of
NGC~3783: the \ovii \ edge and the UTA at $16-17$ \AA \  are
completely blended, making very difficult to isolate the
contribution of the former to the absorption. Our model is
self-consistent in the sense that given the
intrinsic SED of the source, and the column density and ionization
parameter of the absorbing media, the ionization balance is fixed.
By also including the broadening mechanisms of the line as well as 
the rules governing the ion transitions, in a self-consistent way, 
a final global solution consistent with all these processes, can be
obtained. This procedure avoids the inconsistencies that may arise in
the ionization structure of the gas, as well as the effects of
neglecting saturation, that are present in analyses where only an 
ion by ion (or line by line) approach is used to fit the spectral 
features.

The intrinsic continuum and the gas absorption were fitted
simultaneously. Therefore, we iterated many times to get the final
X-ray SED.  We further attenuated the continuum by an equivalent
hydrogen column density of $1.013\times10^{21}$ cm$^{-2}$
\citep{mur96} to account for the Galactic absorption by cold gas
in the direction to the source. In the calculations reported here,
we have explored only photoionization equilibrium models, and we
have assumed solar elemental abundances \citep{grev93}.

\subsection{Continuum \label{con1}}

To model the intrinsic continuum of the source, we first fitted a
simple power law with varying photon index and amplitude. However,
a single power law could not fit the data over the entire range.
We found an excess of flux in the spectrum at energies $\leq 0.6$
keV. This excess could be accounted for with the inclusion of a
black body component superimposed on the power law. Such a thermal
component has been used in the past to describe the NGC~3783
continuum (e.g. De Rosa et al. 2002, and references therein).

To further check our results, we made an independent estimate of
the continuum level using the wavelength bands where no line
features are present, the ``line-free spectral bands'' (horizontal
green ranges at the bottom of Fig.
 \ref{fig1} panels). These bands
were determined excluding the regions with lines in our model,
plus the obvious features that our model missed

\onecolumn

\begin{figure}[b]
\figurenum{3} \plotone{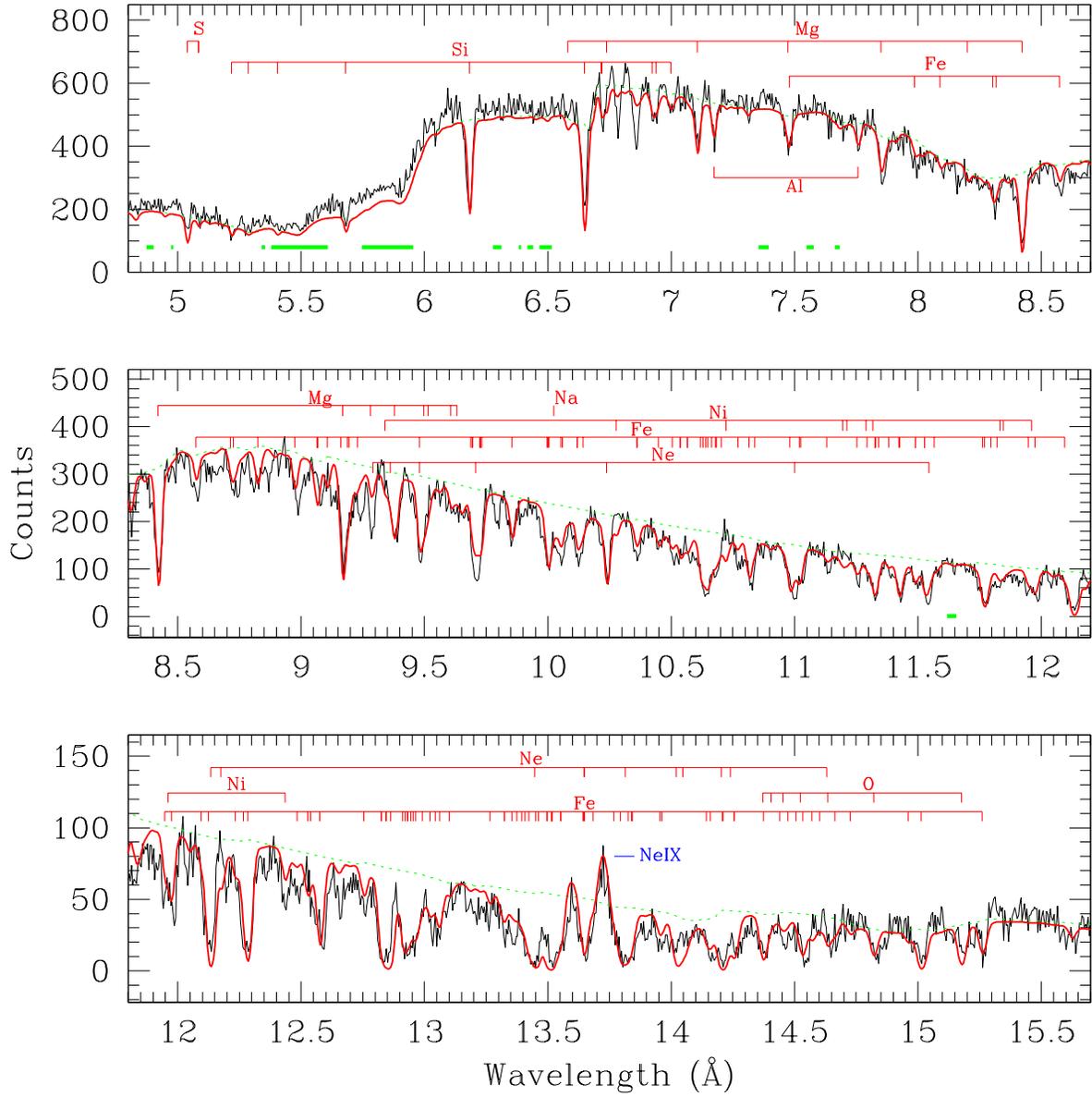} \caption[f3.eps]{ Two phase
absorber model plotted against the first-order MEG spectrum of NGC
3783. Absorption lines predicted are marked in the top (red).
Single labels stand for emission lines (blue). The line free zones
are indicated at the bottom of each panel (green). The continuum
level (including edge continuum absorption) is overplotted for
comparison (dotted green line). The spectrum is presented in the
rest frame system of the absorbing gas.$^9$ \label{fig1}}
\end{figure}

\clearpage

\begin{figure}[b]
\figurenum{3} \plotone{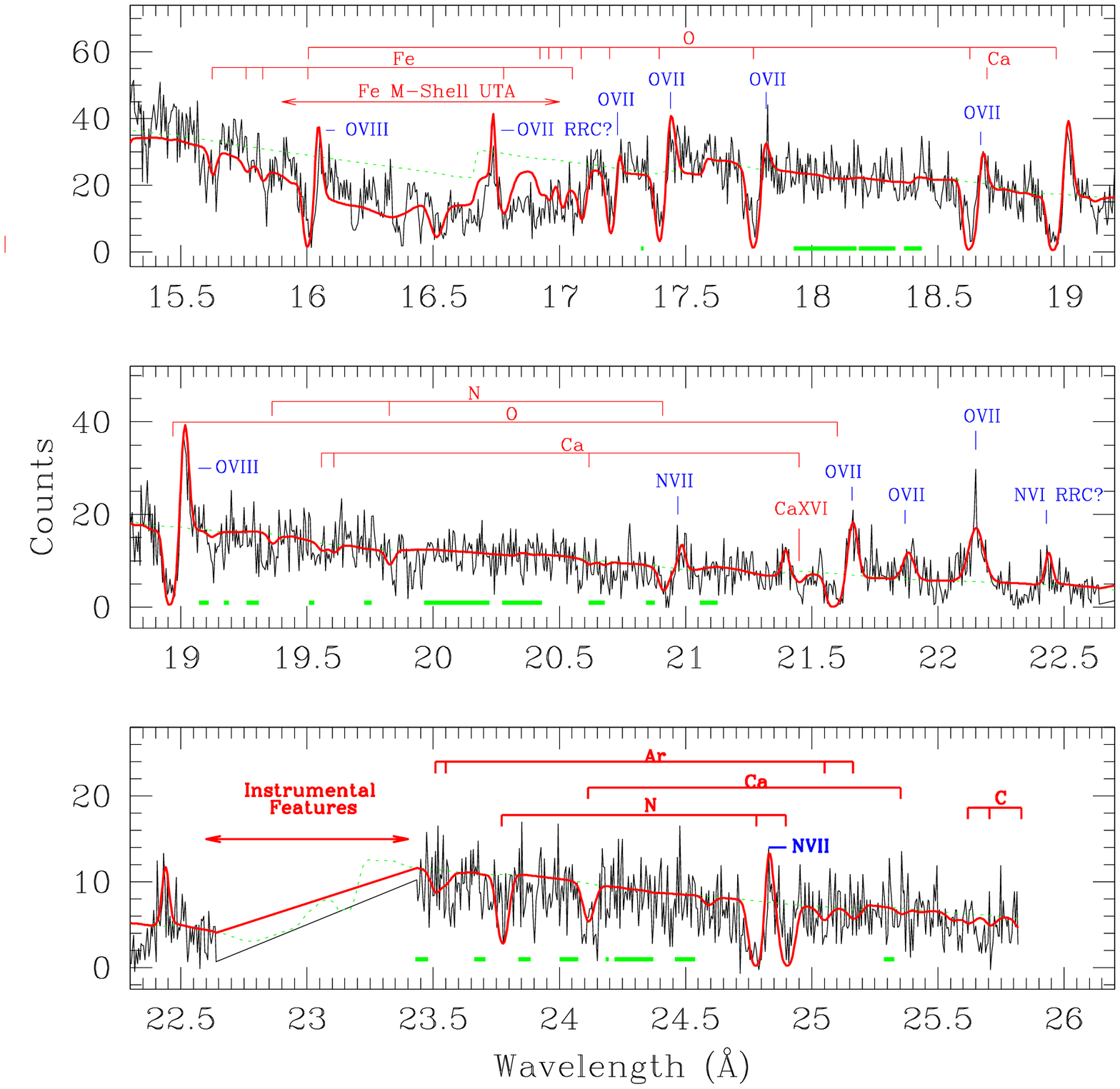} \caption[f4.eps]{Continued}
\end{figure}

\twocolumn

(see \S
\ref{disc}). The two continuum estimates were consistent with each
other. Table \ref{tcont} shows our results. Figure \ref{flux}a
presents the intrinsic continuum of the source, the continuum
attenuated by Galactic absorption, and the observed continuum
(i.e. the continuum further attenuated by the bound-free
transitions from our model). As shown in this figure (and also in
Fig. \ref{fig1}) there is an excess of flux in the observed
continuum at around 5.5-6.5 \AA. We attribute this excess to a
calibration effect due to the rapid changes of the effective area
between 1.7 and 2.3 keV, resulting from the Iridium M-edge. As can
be inferred from the figure, a lower photon index would account
for the difference, but would overpredict the continuum level at
shorter wavelengths.

\begin{figure}[!b]
\figurenum{4} \plotone{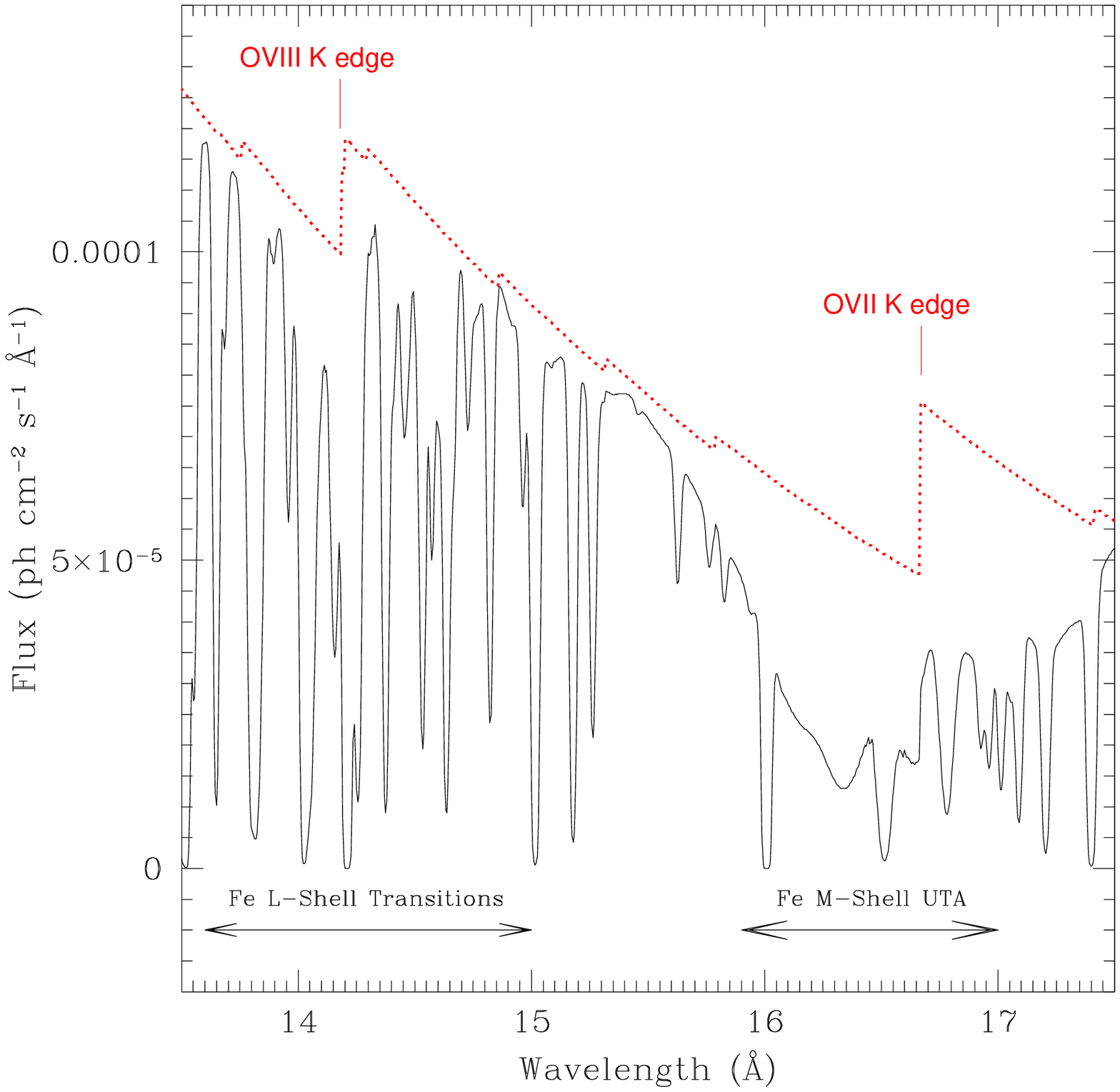} \caption[f5.eps]{PHASE model for
NGC 3783 warm absorber. The top dotted line (red) shows the O
K-edges due to bound free absorption. The lower solid curve
includes additional contribution from bound-bound transitions. As
can be seen, without the inclusion of the resonant absorption, the
optical depth of the edges can be severely overpredicted.$^9$
\label{smita}}
\end{figure}

\begin{figure}[!t]
\figurenum{5} \plotone{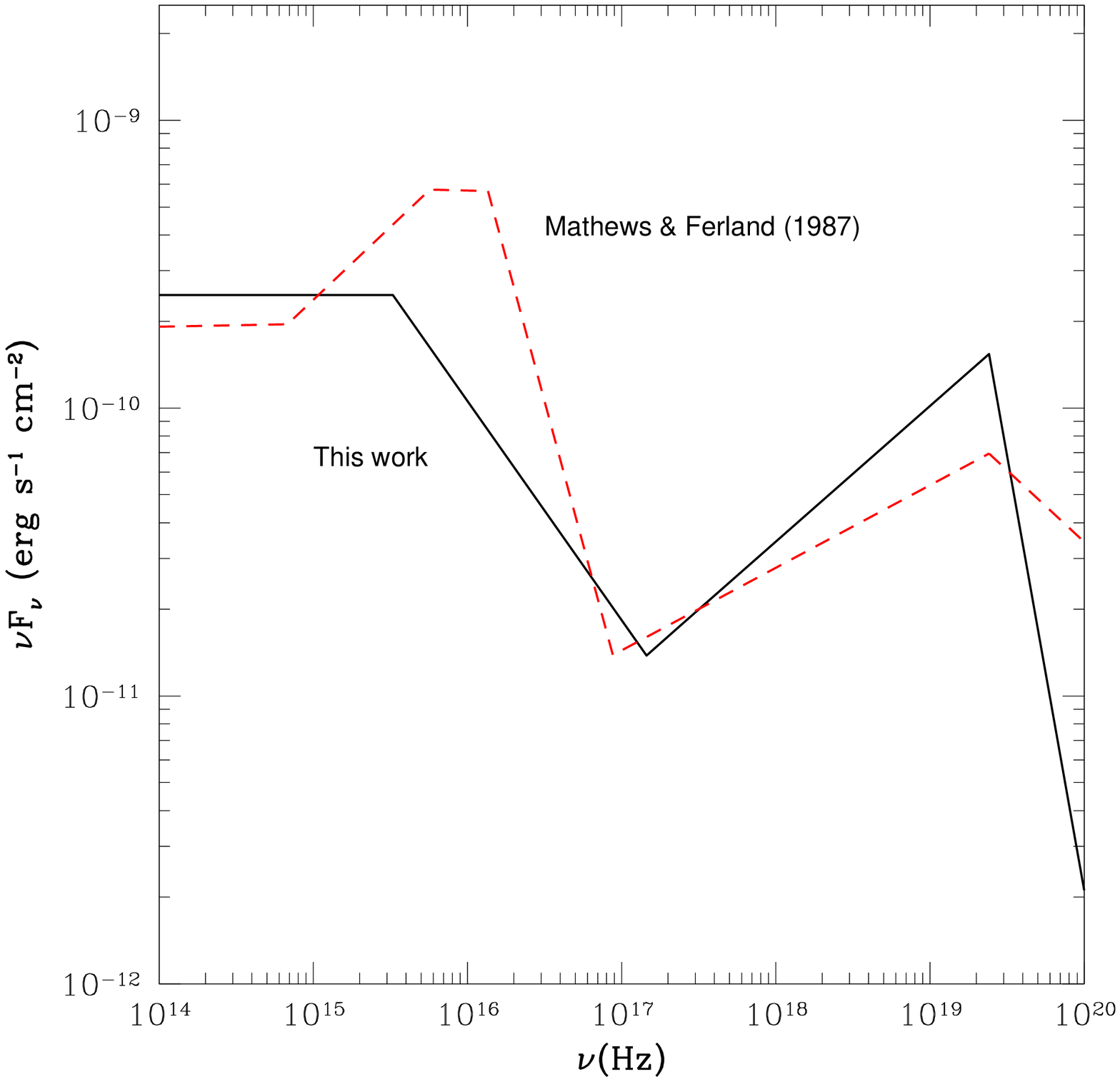} \caption[f6.eps]{Spectral Energy
Distribution used to approximate the NGC 3783 continuum (solid
line). The SED by Mathews \& Ferland (1987) is also plotted for
comparison (dashed line).$^9$  \label{fsed}}
\end{figure}

For simplicity, to calculate the ionization balance, we
approximated the full SED as a power law with the following photon
indices: $\Gamma=2$ below 13.6 eV, $\Gamma=2.4$ in the range 13.6
eV $\leq h\nu\leq$ 0.6 keV, and $\Gamma=1.53$ above 0.6 keV (i.e.
we matched the UV SED derived from UV studies, see below). We
further restricted our SED including high and low energy cutoffs
($\Gamma=5$ above 100 keV and $\Gamma=-3.5$ below 0.1 eV). The SED
is plotted in Figure \ref{fsed}. With the chosen continuum,
between 0.6 keV and 0.1 eV, we assumed the same SED used in
\citet{kraem01} for their modelling of the UV spectrum of NGC
3783. Although this approximation appears reasonable, there are
important uncertainties associated with it, due to the
unobservable region in the UV-X-ray range associated with Galactic
absorption and the lack of simultaneous observations in both
bands. The shape of the SED in the UV and far UV does not affect
the absorption in the X-ray range, as discussed by \citet{ka01}
and Steenbrugge et al. (2003). However, the exact shape of the SED
does affect the values predicted for the ionization parameter, as
will be discussed in \S \ref{sedep}.

\subsection{Emission Features \label{em}}

\begin{deluxetable}{cccc}
\tablecolumns{3} \tablewidth{0pc} \tablecaption{Continuum
Parameters \label{tcont}}

\tablehead{ \multicolumn{3}{c}{Power Law}  } \startdata
Photon Index ($\Gamma$) & Normalization\tablenotemark{a} & $N_{HGal}$ (cm$^{-2}$)\\
1.53$\pm 0.02$ & 0.011$\pm 0.002$ & $1.013\times10^{21}$ \\

\cutinhead{Thermal component}

kT (keV) & Normalization\tablenotemark{b} & \nodata \\

0.10$\pm 0.03$ &  0.00020$\pm 0.00007$ & \nodata \\
\enddata
\tablenotetext{a}{In photons keV$^{-1}$ cm$^{-2}$ s$^{-1}$ at 1
keV.} \tablenotetext{b}{In $L_{39}/D_{10}^2$, where $L_{39}$ is
the source luminosity in units of $10^{39}$ erg s$^{-1}$ and
$D_{10}$ is the distance to the source in units of 10 kpc. }
\end{deluxetable}

\begin{figure}[!t]
\figurenum{6} \plotone{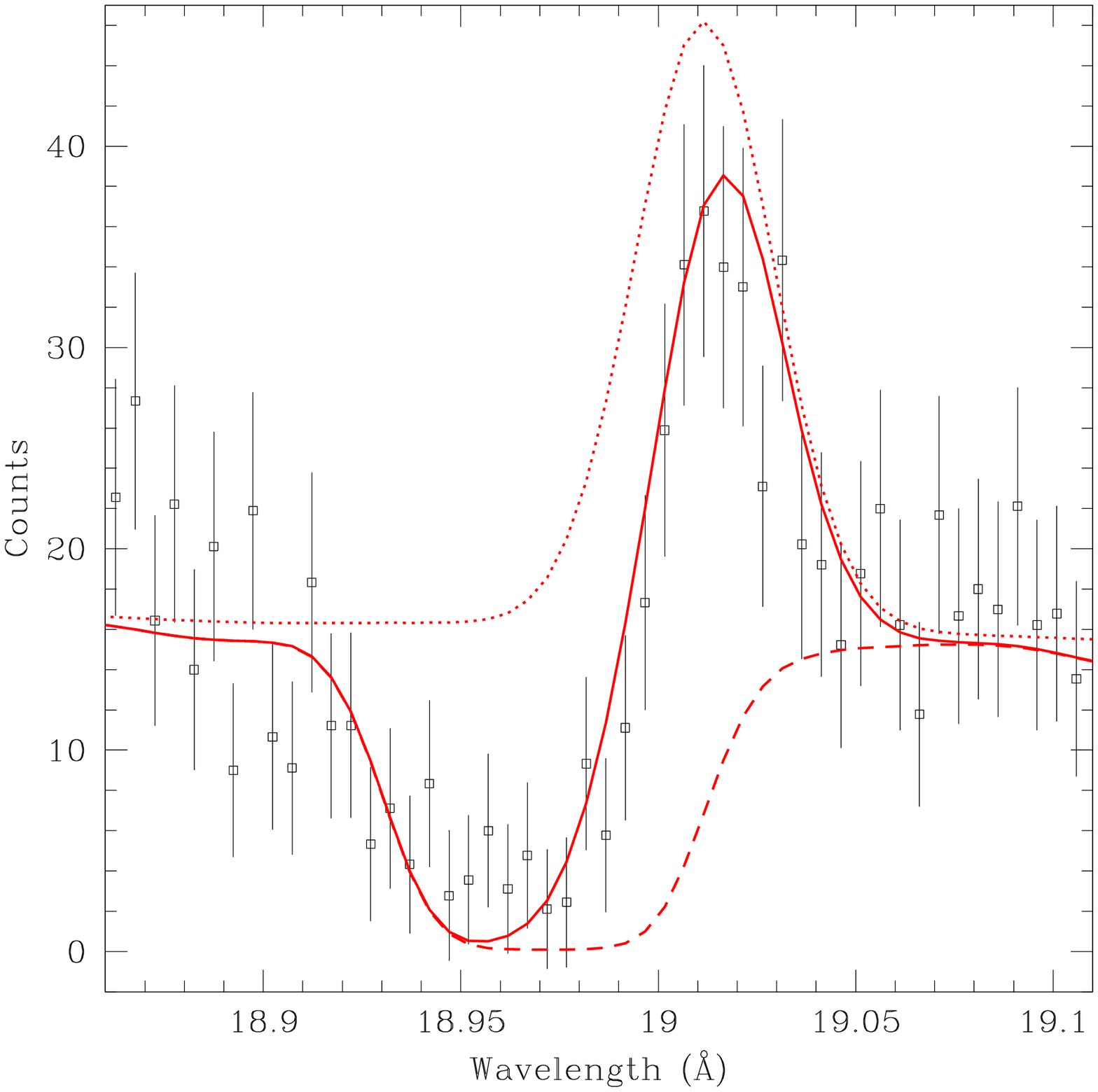} \caption[f7.eps]{
Absorption/emission blending for the O {\sc
viii}($\lambda$18.969\AA) line. While the dotted line stands for
the emission component, the dashed one represents absorption. The
final result (our model) is plotted with the solid line. As can be
observed, the measurement of EWs from the data (e.g. Kaspi et al.
2002) underestimate both absorption and emission.$^9$ \label{ble}
}
\end{figure}

\citet{ka02} identified several emission lines apparent in the 900
ksec spectrum. In our model, these emission lines partly blend
with the absorption lines, such that the depths and EWs of both
appear reduced when the absorption and emission components are
added together (see Fig. \ref{ble}). In its present stage, our
code cannot yet model emission in a self-consistent way. To do so
requires modelling the geometry and kinematics of the absorbing
gas. Therefore, to account for the filling effect of the
absorption lines, we fitted gaussian profiles to the 14 most
prominent emission features in the spectrum, and added them to the
model. To make the fitting as constrained as possible we used the
same outflow velocities and the same FWHMs to model all the
emission lines produced by the same ion. The emission lines fitted
and their properties are listed in Table \ref{eml}. We excluded
the Fe K 6.4 keV complex.

Two features in the spectrum have been tentatively identified as
\ovii \ and N  {\sc vi} radiative recombination continua (RRC),
because of their wavelengths. By contrast to hot collisionally
ionized plasmas where the RRCs are broad extended features, in
photoionized plasmas the RRCs are narrow ``linelike'' features.
The ``width'' of these features can be used, in principle, to
estimate the electron temperature \citep{lie96}.
However, due to the heavy blending of these features with
absorption components, a measurement of the temperature from the
RRCs would be unreliable. To account for this effect, a
self-consistent model for emission processes is also required.

Apart from these two features, all the emission lines have outflow
velocities smaller than those of the absorption components, and
are consistent with being at rest in the system frame (in
accordance with Kaspi et al. 2002). This points to a symmetric
distribution of the emitting gas. Using the \ovii \ triplet plasma
diagnostics by Porquet \& Dubau (2000), an upper limit for the
electron density in the \ovii \ line formation region of
n$_e<2.5\times10^{10}$ cm$^{-3}$ can be imposed, consistent with
the one obtained by Morales et al. (2003, in preparation,
n$_e<2\times10^{10}$ cm$^{-3}$).

\begin{deluxetable}{lcccrl}
\tablecolumns{6} \tablewidth{0pc} \tablecaption{Emission Lines
Parameters  \label{eml}}

\tablehead{ \colhead{Ion} & \colhead{Rest Frame} & \colhead{FWHM}
& \colhead{Outflow Vel.} & \multicolumn{2}{c}{EW$^{a}$} \\
& \colhead{$\lambda$(\AA)} & \colhead{(km s$^{-1}$)} &
\colhead{(km s$^{-1}$)} & \multicolumn{2}{c}{(m\AA)} }

\startdata
Ne{\sc ix}  &   13.699  &      406$\pm169$  &      267$\pm110$      &      15.3& $^{+14.6}_{-9.8} $    \\
O{\sc viii} &   16.006  &      542$\pm343$  &      102$\pm148$      &      25.3& $^{+18.4}_{-16.9} $    \\
RRC O{\sc vii}  &   16.771  &      715$\pm277$  &      1370$\pm173$     &      16.2& $^{+17.6}_{-14.9} $    \\
O{\sc vii}  &   17.200  &      632$\pm269$  &      35$\pm232$   &      7.3& $^{+8.3}_{-8.8} $ \\
O{\sc vii}  &   17.396  &      632$\pm203$  &      35$\pm272$   &      30.3& $^{+12.6}_{-15.2} $    \\
O{\sc vii}  &   17.768  &      632$\pm298$  &      35$\pm150$   &      34.3& $^{+23.4}_{-14.6} $    \\
O{\sc vii}  &   18.627  &      632$\pm222$  &      35$\pm148$   &      25.8& $^{+10.2}_{-17.9} $    \\
O{\sc viii} &   18.969  &      542$\pm235$  &      102$\pm79$       &      79.0& $^{+41.8}_{-34.7} $    \\
N{\sc vii}  &   20.910  &      370$\pm248$  &      198$\pm130$      &      23.7& $^{+33.3}_{-27.1} $     \\
O{\sc vii}  &   21.602  &      632$\pm417$  &      35$\pm208$   &      115.1& $^{+47.4.1}_{-37.5} $   \\
O{\sc vii}  &   21.807  &      632$\pm267$  &      35$\pm265$   &      60.0& $^{+42.5}_{-31.6} $    \\
O{\sc vii}  &   22.101  &      632$\pm258$  &      35$\pm124$   &      144.9& $^{+59.1}_{-68.8} $   \\
RRC N{\sc vi}   &   22.458  &      333$\pm302$  &      1030$\pm106$     &      49.8& $^{+19.4}_{-25.4} $    \\
N{\sc vii}  &   24.781  &      370$\pm281$  &      198$\pm387$      &      43.2& $^{+37.8}_{-41.2} $    \\

\enddata
\tablenotetext{a}{Uncertain due to blending with absorption lines}
\end{deluxetable}

\subsection{Absorption Features}

\begin{figure}[!t]
\figurenum{7} \plotone{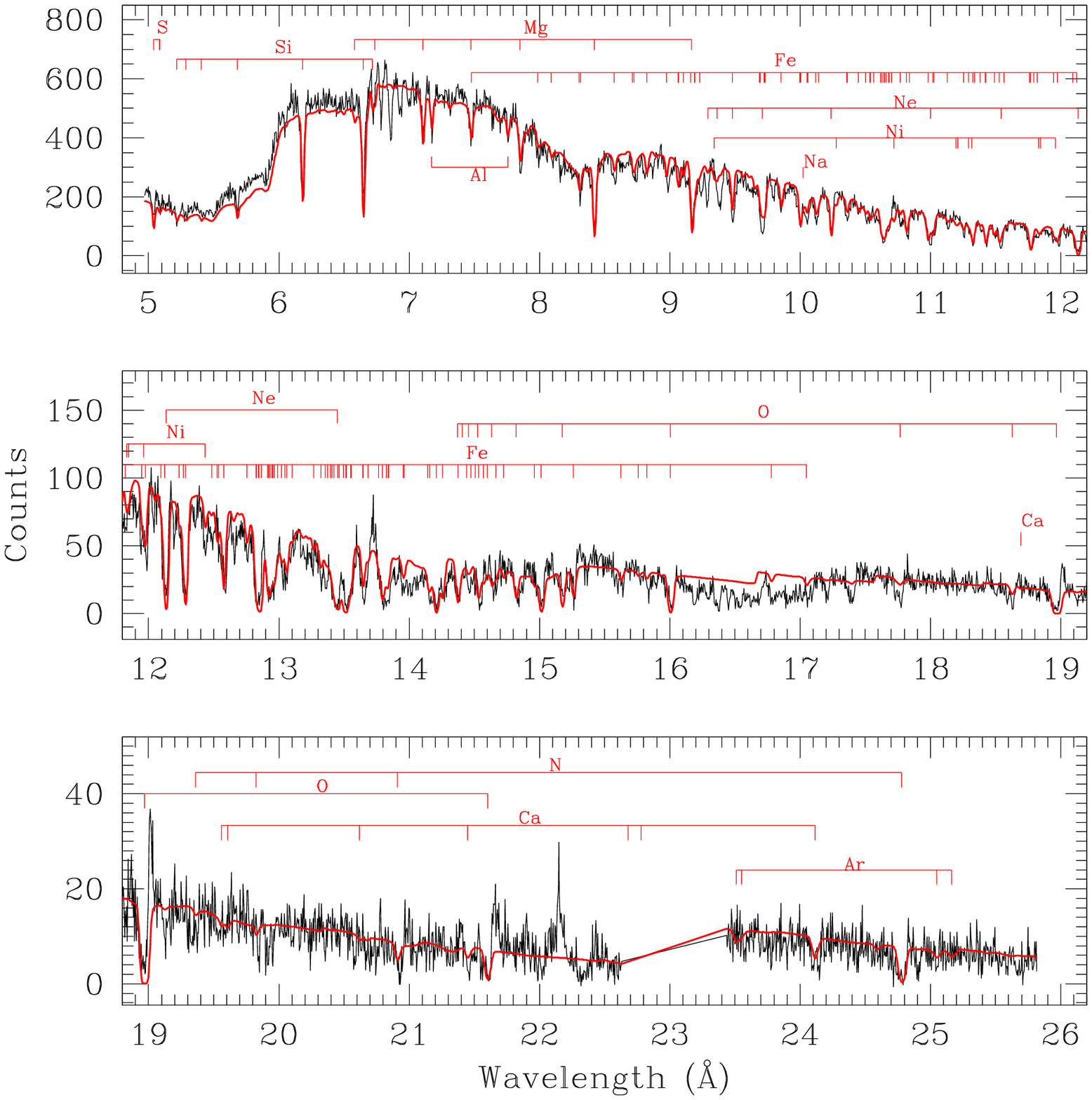} \caption[f8.eps]{ Absorption lines
produced by the high ionization phase (HIP) of our model, plotted
against the first-order MEG spectrum of NGC 3783. The continuum
level contains edge absorption from both components. As can be
observed, below 16 \AA \ the spectrum is dominated by Fe L-shell
absorption lines.$^9$ \label{fig2a} }
\end{figure}

\subsubsection{Components outflowing at 750 km s$^{-1}$ \label{750}}
As was the case for previous studies, the spectrum of NGC 3783
could not be reproduced by a single absorbing component.
Therefore, we included a second one with different ionization.
Both components were consistent with an outflow velocity of
$\approx$750 km s$^{-1}$. Since the absorption lines in the
spectrum are unresolved by the HETGS and many of them are blended,
it was impossible to determine the turbulent velocity of the
systems. Therefore, in order to fit the data, we set this velocity
equal to 300 km s$^{-1}$, the same value used in Kaspi et al.
(2001). The temperature in each gas component was calculated from
our models assuming photoionization equilibrium, and is $\approx
9.5\times10^{5}$ K for the hotter component and $\approx
2.6\times10^{4}$ K for the cooler one; therefore, the contribution
of the thermal velocity to the Doppler velocity is rather small
($<10$ \% for oxygen). Our results are listed in Table \ref{2ab}.
The plot with the model is presented in Figure \ref{fig1} (and
also in Fig. \ref{flux}b, where we present the model for the full
spectral range plotted over the fluxed spectrum).

\begin{figure}[!t]
\figurenum{8} \plotone{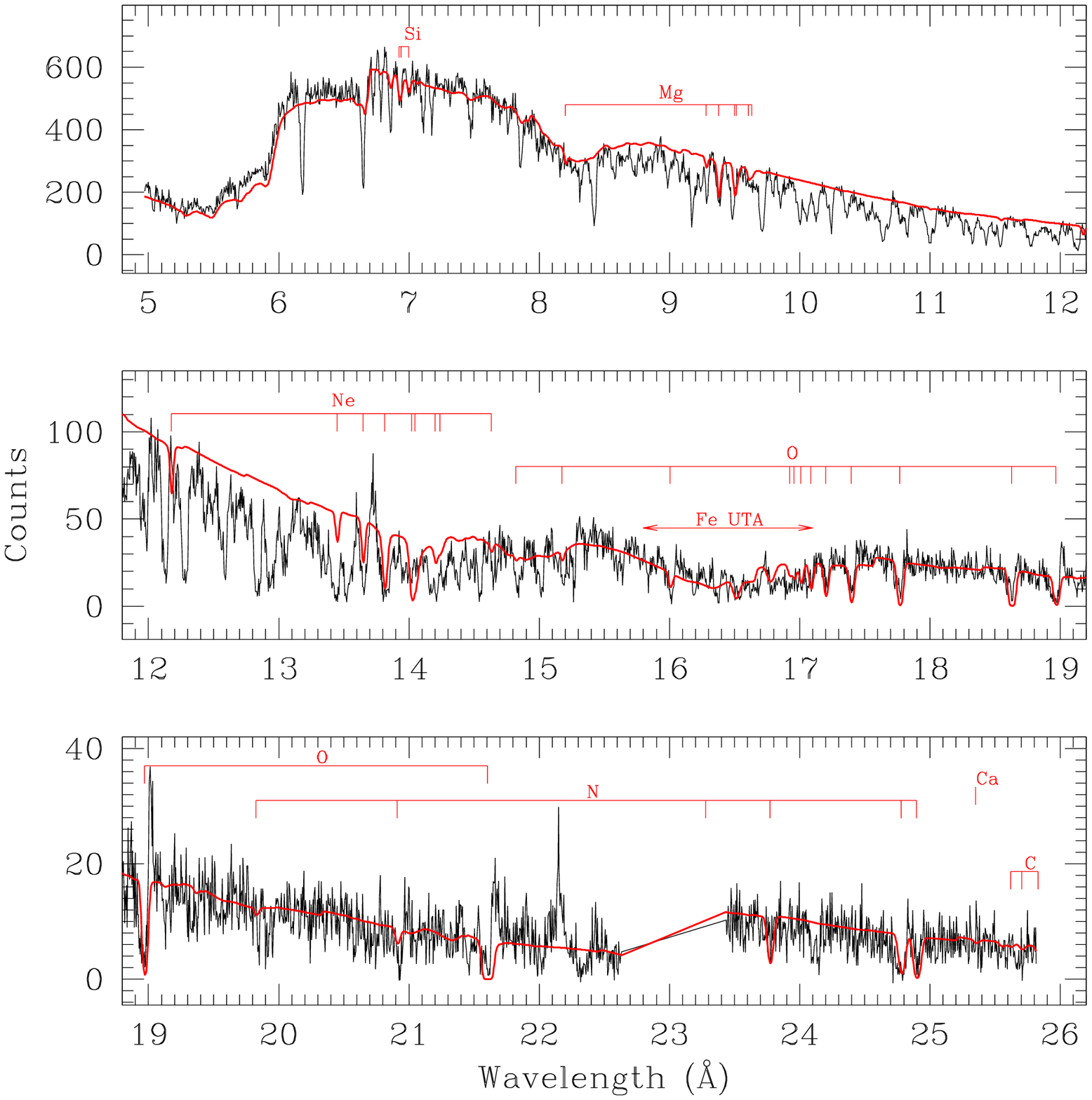} \caption[f9.eps]{ Absorption lines
produced by the low ionization phase (LIP) of our model, plotted
against the first-order MEG spectrum of NGC 3783. The continuum
level contains edge absorption from both components. The main
signature of this phase is the UTA, as well as the \ovii \
lines.$^9$ \label{fig2b}}
\end{figure}

\begin{figure}[!t]
\figurenum{9} \plotone{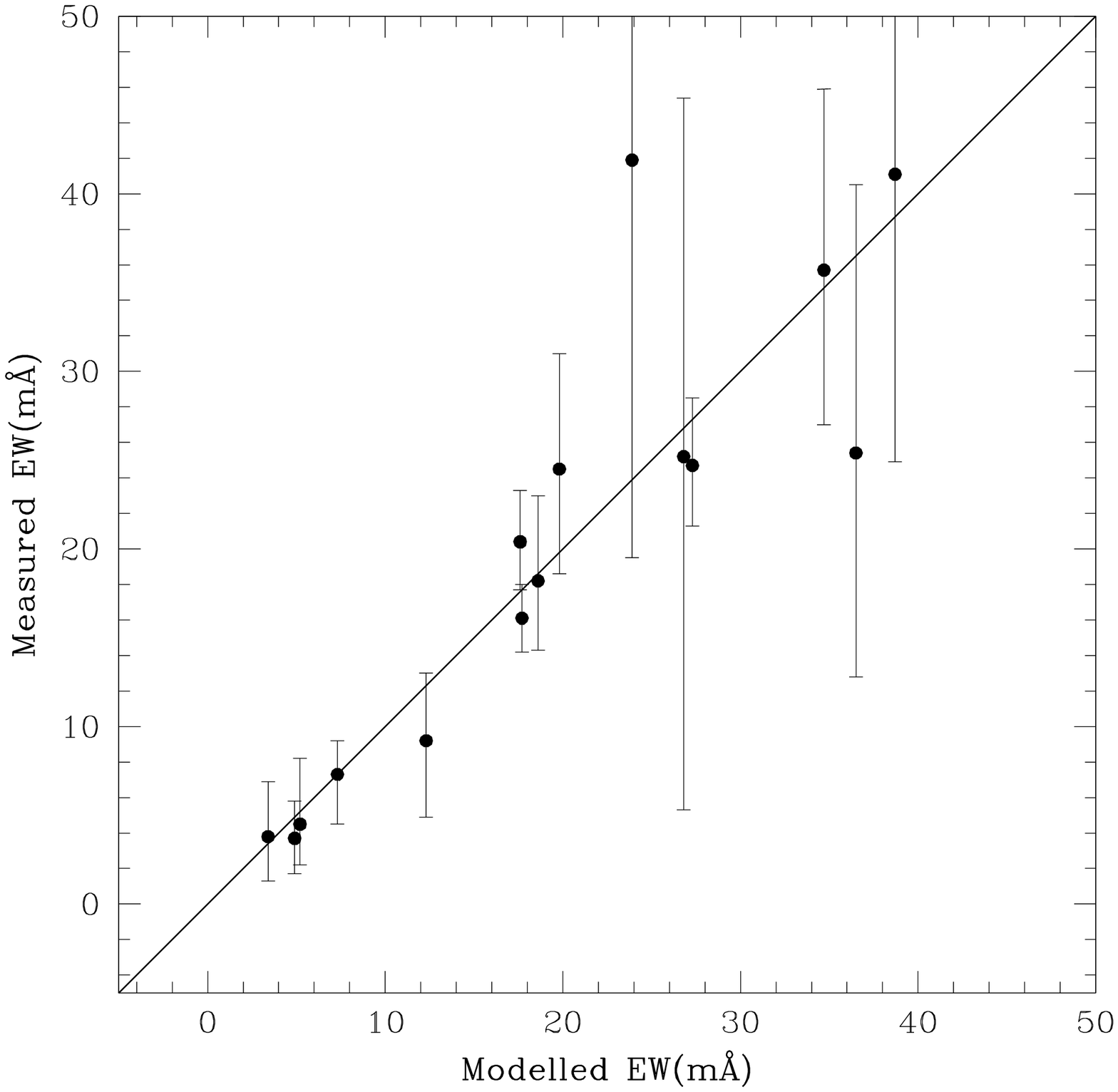} \caption[f10.eps]{ Measured EWs
vs. model predicted EWs for unblended lines in the NGC 3783
spectrum. A line with slope equal to one is presented for
comparison. \label{ew}}
\end{figure}

We find no evidence for cosmic dust or anomalous abundances in any
of the absorbers. It is noteworthy that the outflow of both
components is consistent with a single velocity. While
kinematically indistinguishable and with column densities a factor
4 apart, our two components have dramatically different ionization
parameters: the high ionization phase (HIP) has a value
$\approx35$ times larger than the low ionization phase (LIP). The
HIP, presented in Figure \ref{fig2a}, gives rise to the absorption
by N {\sc vii}, \oviii, Ne {\sc ix-x}, Mg {\sc x-xii}, Al {\sc
xii-xiii}, Si {\sc xi-xiv}, S {\sc xiii-xvi}, Ar {\sc xii-xvii},
Ca {\sc xiii-xviii}, Fe {\sc xvii-xxii}, and Ni {\sc xix-xxi}. On
the other hand, the LIP (presented in Fig. \ref{fig2b}) produces
absorption by C {\sc vi}, N {\sc vi}, O {\sc vii}, Ne {\sc
v-viii}, Mg {\sc vii-ix}, Si {\sc vii-ix}, and the Fe M-shell
inner transitions forming the UTA (Fe {\sc vii-xii}). The most
intense lines, as well as the unblended features, are listed in
Table \ref{abl}, where we show the contribution from the HIP and
the LIP separately. All the EWs derived from our model for
unblended lines agree with those measured from the data (see Fig.
\ref{ew}).

\begin{deluxetable}{lcc}
\tablecolumns{3} \tablewidth{0pc} \tablecaption{Two Phase Absorber
Parameters
 \label{2ab}}

\tablehead{\colhead{Parameter} & \colhead{High-Ionization} &
\colhead{Low-Ionization} } \startdata
Log U$^a$ &  0.76$\pm0.1$ & -0.78$\pm0.13$ \\
Log N$_{H}$ (cm$^{-2}$)$^a$ &  22.20$\pm.22$ & 21.61$\pm0.14$ \\
V$_{Turb}$ (km s$^{-1}$)& 300 & 300 \\
V$_{Out}$ (km s$^{-1}$)$^a$& 788$\pm138$ & 750$\pm138$ \\
T (K)$^b$& $9.52\pm0.44\times10^5$ & $2.58\pm0.39\times10^4$ \\
$[$Log T (K)$]$& 5.98$\pm0.02$ & 4.41$\pm0.07$ \\
Log T/U ($\propto$P$^c$)& 5.22$\pm0.12$ & 5.19$\pm0.20$ \\
Log $\Xi$$^d$ & 1.02$\pm0.12$ & 0.99$\pm0.20$ \\
\enddata
\tablenotetext{a}{Free parameters of the model.}
\tablenotetext{b}{Derived from the column density and ionization
parameter, assuming photoionization equilibrium.}
\tablenotetext{c}{The pressure P$\propto$n$_e$T. Assuming that
both phases lie at the same distance from the central source
n$_e\propto$1/U, and P$\propto$T/U.} \tablenotetext{d}{Ionization
Parameter as in Krolik \& Kallman 1984), i.e. the ratio between
the radiation pressure and the gas pressure.}
\end{deluxetable}

\subsubsection{Component outflowing at 1345 km s$^{-1}$ \label{hivel}}
In the 900 ksec spectrum Kaspi et al. (2002) found evidence for two X-ray absorbers in the
lines of \ovii, but could not conclude the same in those of Ne
{\sc x}.  We also see evidence for these
\ovii \ lines (lower panels of Fig. \ref{figuv}). As can be
inferred from Table \ref{abl}, while our model indicates that all
the absorption produced by Ne {\sc x} arises in the HIP, most of
the \ovii \ absorption originates in the LIP. Furthermore, no
evidence of a second absorber in \oviii \ is found, and this ion
too is mainly produced by the HIP (Table \ref{abl}). Therefore,
the detection of these \ovii \ features indicates the presence of
a low ionization, high velocity component. According to our models
(see \S \ref{uv} and Table \ref{uva1}) a component capable of
reproducing the absorption lines in \ovii \, but no absorption in
\oviii \ or Ne {\sc x} at 1365 km s$^{-1}$, would need a low
ionization parameter, but also a low column density (log
N$_H\sim$ 21) and therefore, would only produce observable
absorption features by Ne {\sc vi-vii} between 13.5 and 14 \AA
($1s-2p$ transitions mainly), and N {\sc vi} between 23.1 and 23.8
\AA \ and at 24.9 \AA ($1s-2p$ and $1s-3p$ transitions). Since the
Ne {\sc vi-vii} lines are weak and lie in a region dominated
by the Fe L-shell transitions, they are undetectable. The N {\sc
vi} lines are also undetectable due to low signal to noise ratio
and calibration uncertainties at long wavelengths. Therefore, the \ovii
\ lines are really the only features observable from this
component. This component would also produce
absorption by O{ {\sc vi}, and an absorber at roughly this outflow
velocity has been already clearly observed in the UV (see \S
\ref{uv}). This UV detection gives us greater confidence in the
reality of the high velocity LIP absorber, which is otherwise of
borderline significance in the X-ray data.

\subsection{UV Absorbers \label{uv}}

\begin{figure}[!t]
\figurenum{10} \plotone{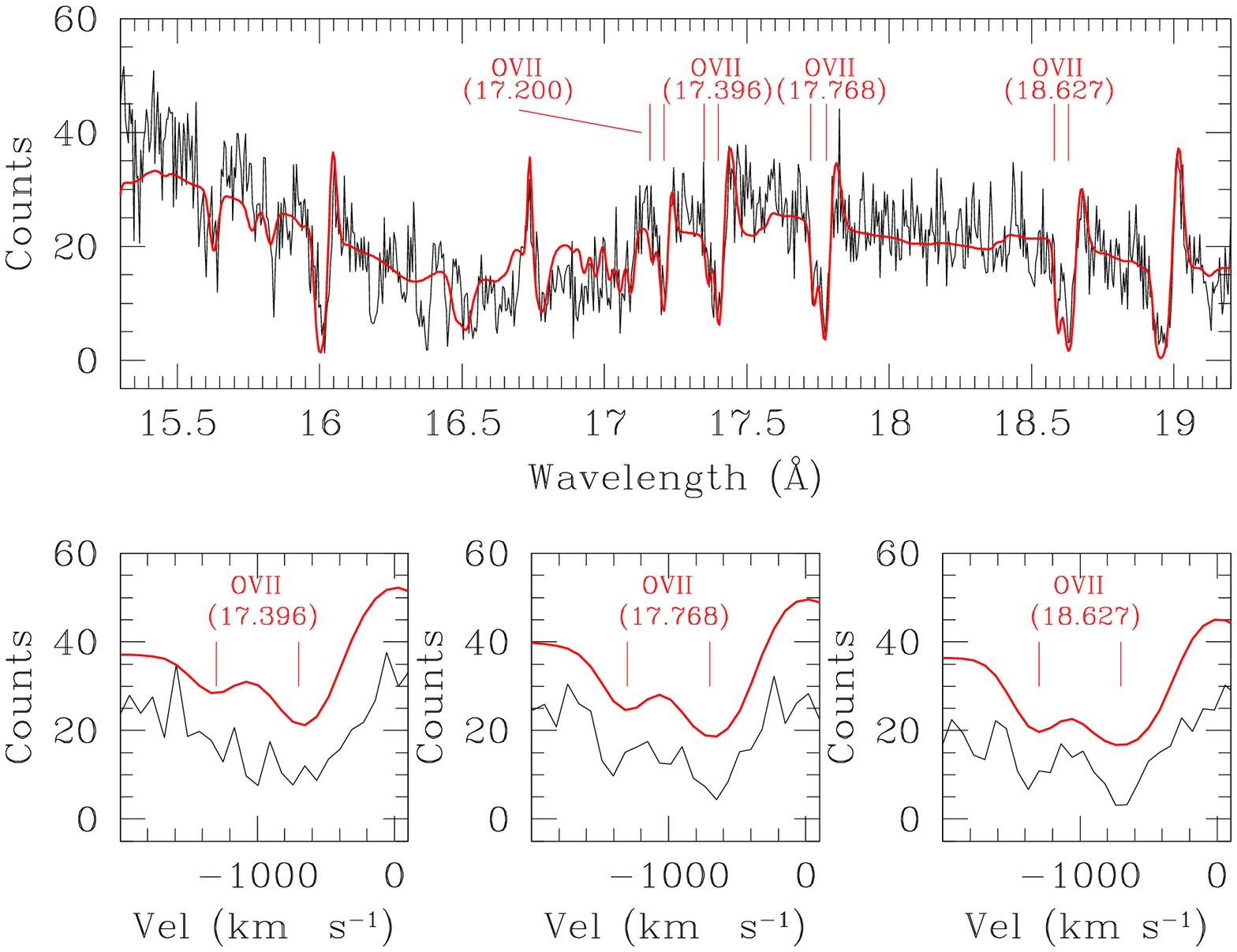} \caption[f11.eps]{Low ionization
phase decomposed in three subcomponents matching the FWHM, outflow
velocity, and H column density of the absorbers predicted in the
UV band (Kraemer et al. 2001). The ionization parameters, however,
are different (see \S \ref{uv}). The model is presented only in
the 15.5-19 \AA \ range. In other ranges, there is no significant
difference with the model presented in Fig. \ref{fig1}. As can be
observed, the two observed components of \ovii \ are predicted.
The bottom panels show the 3 most intense \ovii \ lines in
velocity space. The model has been shifted up 10 counts for
clarity.$^9$ \label{figuv}}
\end{figure}

Kraemer, Crenshaw \& Gabel (2001) and Gabel et al. (2003) have
presented the UV spectrum of NGC 3783. Absorption lines due to
\civ, \nv, and \ovi \ are clearly present in the HST and FUSE
spectra, with each line showing four different components
corresponding to four outflow velocities. One of these components
(outflow $\approx$ 1027 km s$^{-1}$) is too weak to be detected
with X-ray instruments (see Fig. \ref{fab}). However, the other
three components are consistent with the outflow velocities of the
X-ray LIP \ovii \ absorbers (at 750 km s$^{-1}$ and the 4 \ovii \
lines at 1345 km s$^{-1}$). Furthermore, the ionization degree of
the LIP X-ray components can also produce UV features (in contrast
with the HIP, which is almost transparent in the UV band due to
its high ionization degree). Motivated by this, and with the aim
of exploring the connection between UV and X-ray absorbers, we
decided to model the X-ray data for comparison with the models by
Kraemer et al. (2001).

To begin with, we considered model parameters exactly as in
Kraemer et al. (2001): the total equivalent column density N$_H$,
ionization parameter U, three different velocity components for
each line, with FWHM and outflow velocity V$_{out}$ as for the UV
lines (see Table \ref{uva1}). We found that the predicted
strengths of X-ray lines do not match the data (see Fig.
\ref{si}a). Then we changed only one input parameter, the
ionization parameter U. We fitted the X-ray spectrum, leaving U as
a free parameter, but with all other parameters as given above. A
good fit to the data was obtained, as shown in figure \ref{figuv},
for the 15.5 to 19.0 \AA\ range (in other ranges, there is no
significant difference with respect to the model presented in
Figure \ref{fig1}, as discussed in \S \ref{hivel}). However, the
best fit value of U is smaller by at least a factor of 3, than
that found by Kraemer et al. for the UV spectrum (Table
\ref{uva1}). In principle, this can be an effect of the different
input continuum that we used, so we repeated the analysis by using
exactly the same SED as in Kraemer et al. (i.e. with a value of
$\Gamma=1.8$) above 0.6 keV and no high/low energy cutoffs). The
resulting U still did not match the Kraemer et al. value. The
degree of ionization found by Kraemer et al. is too high to fit
the X-ray absorber. A na\"{\i}ve interpretation of these facts
could be that the UV and X-ray absorbers are two different
phenomena, independent of each other. This, however, is not true.
As discussed below in \S5, the high values of U inferred from the
UV data alone, would produce strong absorption by the UTA in the
region around 16 \AA\, and this is not consistent with the
observations. Figure \ref{si}a illustrates this point clearly.

The excellent level of agreement between the data and our new
model with three different velocity components for each line (as
in the UV) is noteworthy. This strongly indicates a common origin
of UV and X-ray absorbers, as we have been suggesting for years
(Mathur et al. 1994, 1995, 1997, 1998, 1999; Monier et al.

\begin{deluxetable}{lccrrrl}
\tablecolumns{7} \tablewidth{0pc} \tablecaption{Most intense
unblended absorption lines \label{abl}}

\tablehead{ & &  \multicolumn{4}{c}{Equivalent Width (m\AA)} \\
\cline{3-7}
\colhead{Ion} & \colhead{Wavelength (\AA)} & \colhead{HIP} &
\colhead{LIP} & \colhead{Total} & \multicolumn{2}{c}{Measured} }
\startdata
Si{\sc xiv}     &   6.182   &   16.1    &   $<0.5$ &   16.1    &   17.7&$^{+1.9}_{-1.9}$    \\
Si{\sc xiii}    &   6.648   &   20.4    &   $<0.5$ &   20.4    &
17.6&$^{+2.9}_{-2.7}$   \\
Mg{\sc xii}     &   7.106   &   7.3 &   $<0.5$ &   7.3 &   7.3&$^{+1.9}_{-2.8}$ \\
Al{\sc xiii}    &   7.173   &   3.7 &   $<0.5$ &   3.7 &   4.9&$^{+2.1}_{-2.0}$ \\
Al{\sc xii}     &   7.757   &   3.8 &   $<0.5$ &   3.8 &
3.4&$^{+3.1}_{-2.5}$  \\
Fe{\sc xxiii}       &   8.304   &   5.1 &   $<0.5$ &   5.1 &
11.7$^c$&$^{+4.3}_{-4.1}$     \\
Mg{\sc xii}     &   8.421   &   24.7    &   $<0.5$ &   24.7    &   27.3&$^{+3.8}_{-3.4}$    \\
Fe{\sc xxii}    &   8.714   &   4.5 &   $<0.5$ &   4.5 &   5.2&$^{+3.7}_{-2.3}$ \\
Ne{\sc x}       &   10.239  &   20.7    &   $<0.5$ &   20.7    &
23.0$^c$&$^{+3.8}_{-3.2}$ \\
Fe{\sc xix}     &   10.816  &   18.2    &   $<0.5$ &   18.2    &   18.6&$^{+4.8}_{-3.9}$    \\
Fe{\sc xxii}    &   11.770  &   28.4    &   $<0.5$ &   28.4    &
41.0$^c$ &$^{+8.8}_{-7.4}$   \\
Ne{\sc x}       &   12.134  &   45.4    &   $<0.5$ &   45.4    &   $^a$ &   \\
Fe{\sc xxi}     &   12.284  &   39.7    &   $<0.5$ &   39.7    &   47.4$^c$&$^{+8.3}_{-7.1}$    \\
Fe{\sc xx}      &   12.576  &   35.7    &   $<0.5$ &   35.7    &   34.7&$^{+10.2}_{-8.7}$    \\
Fe{\sc xx}      &   12.754  &   9.2 &   $<0.5$ &   9.2 &   12.3&$^{+3.8}_{-4.3}$    \\
Fe{\sc xx}      &   12.824  &   26.1    &   $<0.5$ &   26.1    &   $^a$ &   \\
Fe{\sc xx}      &   12.846  &   38.9    &   $<0.5$ &   38.9    &   $^a$ &   \\
Fe{\sc xx}      &   12.864  &   37.0    &   $<0.5$ &   37.0    &   $^a$ &   \\
Fe{\sc xix}     &   13.423  &   26.1    &   $<0.5$ &   26.1    &   $^a$ &   \\
Ne{\sc ix}      &   13.447  &   31.5    &   11.0    &   42.6    &   $^a$ &   \\
Fe{\sc xix}     &   13.462  &   30.0    &   $<0.5$ &   30.0    &   $^a$  &  \\
Fe{\sc xix}     &   13.497  &   36.4    &   $<0.5$ &   36.4    &   $^a$  &  \\
Fe{\sc xix}     &   13.518  &   48.3    &   $<0.5$ &   48.3    &   $^a$  &  \\
Fe{\sc xix}     &   13.795  &   28.5    &   $<0.5$ &   28.5    &   $^a$  &  \\
Ne{\sc vii}     &   13.814  &   $<0.5$ &   42.3    &   42.3    &   $^a$  &  \\
Ne{\sc vi}      &   14.020  &   $<0.5$ &   41.1    &   41.1    &   $^a$  &  \\
Ne{\sc vi}      &   14.047  &   $<0.5$ &   51.4    &   51.4    &   $^a$  &  \\
Fe{\sc xviii}       &   14.208  &   49.2    &   $<0.5$ &   49.2    &   $^a$ &   \\
Fe{\sc xviii}       &   14.208  &   41.4    &   $<0.5$ &   41.4    &   $^a$ &   \\
Fe{\sc xviii}       &   14.373  &   30.6    &   $<0.5$ &   30.6    &   $^a$ &   \\
O{\sc viii}     &   14.821  &   22.3    &   3.1 &   25.4    &   36.5&$^{+15.1}_{-12.6}$   \\
Fe{\sc xix}     &   14.961  &   9.8 &   $<0.5$ &   9.8 &
9.5$^c$&$^{+6.4}_{-5.6}$ \\
Fe{\sc xvii}    &   15.014  &   54.6    &   $<0.5$ &   54.6    &
33.7$^c$&$^{+17.6}_{-15.2}$ \\
O{\sc viii}     &   15.176  &   34.6    &   6.5 &   41.1    &   38.7&$^{+14.4}_{-16.2}$   \\
Fe{\sc xvii}    &   15.261  &   24.5    &   $<0.5$ &   24.5    &   19.8&$^{+6.5}_{-5.9}$    \\
O{\sc viii}     &   16.006  &   51.0    &   16.7    &   67.7    &
59.8$^b$&$^{+10.8}_{-17.4}$ \\
O{\sc vii}  &   17.086  &   $<0.5$ &   24.5    &   24.5    &   $^b$ &   \\
O{\sc vii}  &   17.200  &   $<0.5$ &   33.0    &   33.0    &
28.9$^b$&$^{+6.8}_{-7.6}$ \\
O{\sc vii}  &   17.396  &   $<0.5$ &   43.4    &   43.4    &
38.9$^b$&$^{+7.8}_{-11.2}$ \\
O{\sc vii}  &   17.768  &   5.0 &   55.2    &   60.2    &   40.1$^b$&$^{+16.0}_{-15.2}$ \\
O{\sc vii}  &   18.627  &   13.2    &   71.0    &   84.2    &
46.3$^b$&$^{+16.9}_{-14.2}$ \\
O{\sc viii}     &   18.969  &   89.0    &   54.1    &   143.1   &
55.0$^b$&$^{+23.8}_{-15.4}$ \\
N{\sc vii}  &   20.910  &   19.3    &   12.8    &   32.1    &29.5$^b$&$^{+23.3}_{-21.4}$   \\
O{\sc vii}  &   21.602  &   49.5    &   128.3   &   177.8   &34.2$^b$&$^{+30.8}_{-33.1}$   \\
N{\sc vi}   &   23.277  &   $<0.5$ &   24.8    &   24.8    &   $^a$ &   \\
N{\sc vi}   &   23.771  &   $<0.5$ &   41.9    &   41.9    &23.9& $^{+21.3}_{-22.5}$    \\
Ca{\sc xiv}     &   24.114  &   25.2    &   $<0.5$ &   25.2
&26.8& $^{+20.2}_{-19.9}$   \\
N{\sc vii}  &   24.781  &   67.0    &   54.9    &   121.9   &
66.5$^b$
&$^{+37.2}_{-34.4}$    \\
N{\sc vi}   &   24.898  &   $<0.5$ &   68.7    &   68.7    &33.9$^b$&$^{+31.7}_{-27.2}$  \\

\enddata
\tablenotetext{a}{Not measured due to heavy blending with several
absorption lines} \tablenotetext{b}{EWs do not take into account
blending with the corresponding emission line}
\tablenotetext{c}{EWs do not take into account blending with other
absorption line}

\end{deluxetable}

\clearpage

\begin{deluxetable}{lcccc}
\tablecolumns{4} \tablewidth{0pc} \tablecaption{Parameters found
for U, assuming UV absorber values for N$_H$, FWHM, and V$_{Out}$.
\label{uva1}}

\tablehead{
 \colhead{Parameter} & & \colhead{Comp 1} &
\colhead{Comp 2} & \colhead{Comp 3}  } \startdata
Log N$_{H}$ (cm$^{-2}$)$^a$ & fix& 21.08 & 20.81 & 21.18  \\
FWHM (km s$^{-1}$)$^a$& fix& 193 & 170 & 280  \\
V$_{Out}$ (km s$^{-1}$)$^a$& fix& 1365 & 548 & 724 \\
Log U(UV)$^a$ & \nodata&  -0.11 &  -0.1 &  -0.19  \\
Log U(X-ray)$^b$ & free &  -1.20$\pm0.09$ & -0.69$\pm0.21$  &
-0.99$\pm0.11$
\enddata
\tablenotetext{a}{Values predicted in the UV by Kraemer et al.
2001.} \tablenotetext{b}{The only free parameter in this model.
Calculated assuming N$_H$, FWHM, and V$_{Out}$ values from the
UV.}
\end{deluxetable}

2001, although with a one phase model; only high resolution spectra
allows the identification of two components). The above exercise also
highlights the danger of modelling the absorbers using UV data
alone. The UV value of the ionization parameter underpredicts the
ionic column densities by as much as factors of ten. The most
likely explanation of this discrepancy is that the UV lines are
saturated and the saturation effects are often estimated
inadequately. This is illustrated beautifully in the case of NGC
5548 by Arav et al. \ (2003). They find that the \ovi \ column
density measured from UV data by Brotherton et al. (2002) is lower
than the robust lower limit obtained from the XMM-Newton data, by
at least a factor of seven. When the effects of saturation are
dealt with in a more sophisticated way by Arav et al. a good
agreement is found.

Saturation was the main point in the models of Mathur et al. who
parameterized its effect using the simple technique of
curve-of-growth analysis. The key point that we stress again here
is that the X-ray data imply a large amount of UV absorption,
which cannot be hidden by claiming that the UV and X-ray absorbers
are different. We will discuss the relationship between the UV and
X-ray absorption in NGC 3783 in a forthcoming paper.

\begin{figure}[!b]
\figurenum{11} \plotone{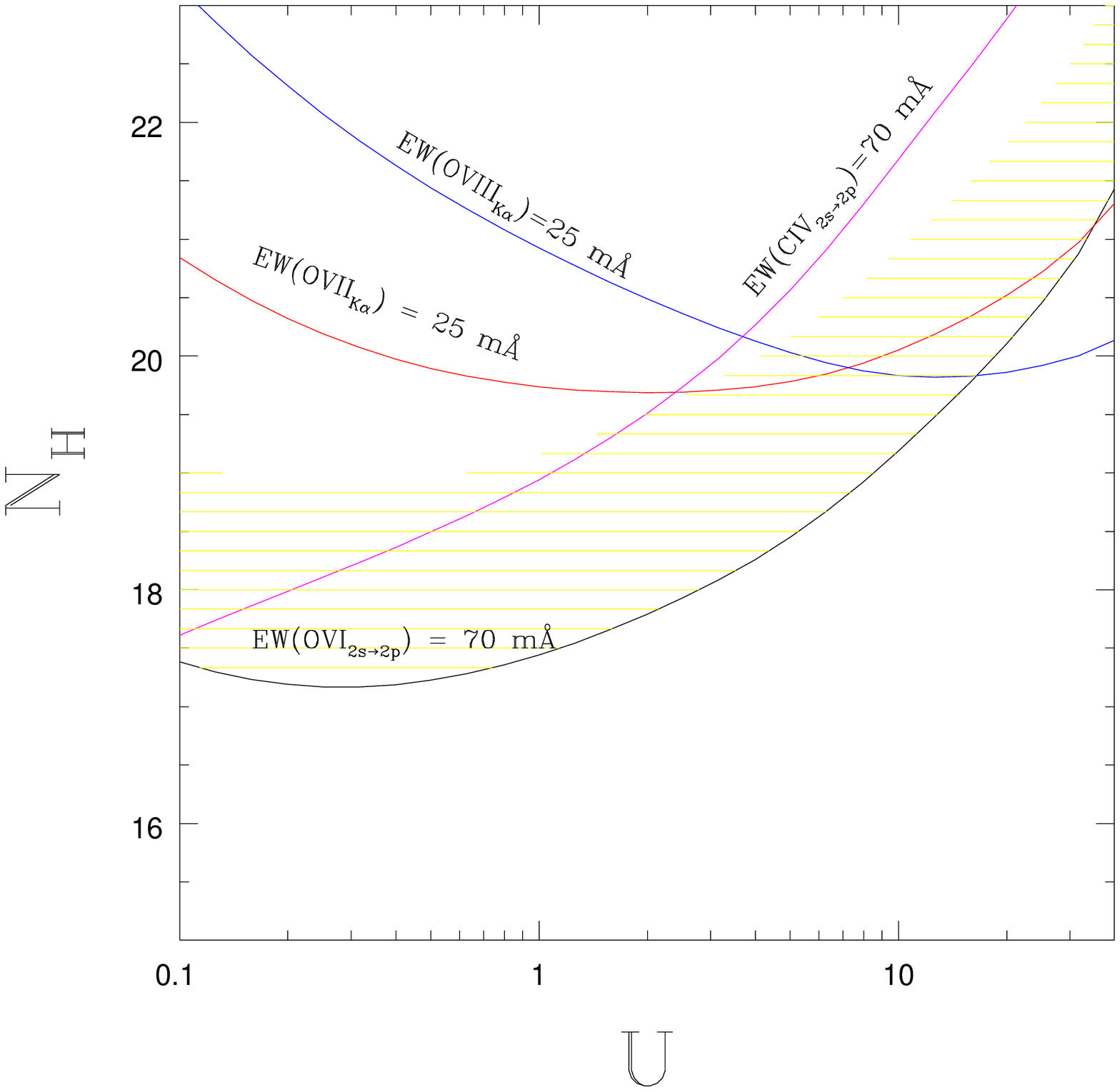} \caption[f12.eps]{Equivalent
hydrogen column density (N$_H$) vs. ionization parameter ($U =
n_{phot}/n_e$), showing observational lower limits for two UV
lines (to a 3$\sigma$ level): CIV (HST/STIS), OVI (FUSE), and two
X-ray lines: OVII, OVIII ({\em Chandra/LETG}). The limits are
calculated for lines of width, $b$=100~km~s$^{-1}$, a UV flux
f(1032\AA)=5$\times$10$^{-14}$erg~cm$^{-2}$s$^{-1}$, 20~ksec HST
and 50~ksec FUSE exposures; an X-ray flux of f(0.5-2~keV)=2~mCrab,
and a 100~ksec LETG exposure.  The shaded region shows where the
OVI measurement is unsaturated (i.e. lies on the linear part of
the curve of growth).$^9$ \label{fab} }
\end{figure}

\section{Evaluation of the Spectral Fitting Results \label{disc}}

The self-consistent absorption model we present in this paper is
the first to be carried out for the 900 ksec spectrum of NGC 3783.
The extraordinary signal to noise ratio of the data has allowed us
to obtain important constraints that make the modelling results
more reliable. This is an important point since without the
restrictions imposed by the data (see below), small changes in the
fitting can give rise to completely different physical
interpretations of the medium (see \S \ref{comp}).

Our model presents a surprisingly simple picture: two
kinematically indistinguishable gas outflows with quite
different photoionization equilibrium temperatures and two widely
separated ionization parameter values. The model also suggests
that no third component with high H column density and
intermediate U is consistent with the data (see below). Six free
parameters are sufficient to model $>$ 100 absorption features and
to correctly predict the EW of the 23 most intense lines that have
little or no blending. Both components are consistent with solar
abundances. Our fits are sensitive to departures of the Mg/O,
Fe/O, Fe/Mg, Ne/O ratios by any more than a factor of $\sim$2 from
solar, since transitions from these elements are in regions of
high S/N. These data are not sensitive, however, to the C/N
relative abundance. A component with low U and N$_H$ at 1345 km
s$^{-1}$ is needed to model an asymmetry in 4 \ovii \ lines. A
fourth component is evidenced by the presence of two Fe {\sc xxv}
absorption lines at 1.57 and 1.85 \AA \ (identified by Kaspi et
al. 2002). However, such component is even more highly ionized, and
only contributes with a few lines to the overall spectrum. Almost
all these features are below 4 \AA, and are only detected by the
HEG configuration of the HETG (not modelled here). This component
is not well constrained (particularly by the MEG band), and we defer
this analysis to a forthcoming paper. 

Absorption by iron and oxygen is particularly important in the
spectrum of NGC 3783. Even though some of their features are
blended, these elements can be regarded as the main ``tracers'' of
the gas, and therefore, can be used to evaluate the validity of a
given model. In our case, an important result is that while \oviii
\ and Fe {\sc xvii-xxii} are the main features of the hotter phase
(see Fig. \ref{fig2a}), the signature of the cooler phase can be
attributed to the UTA and the \ovii \ lines (see Fig.
\ref{fig2b}). Only 2\% of the \ovii, but 90\% of the \oviii,
arises in the HIP (N$_{LIP}$(\ovii)$\approx 1.5 \times 10^{18}$
cm$^{-2}$ and N$_{LIP}$(\oviii)$\approx 1.2 \times 10^{17}$
 cm$^{-2}$ vs. N$_{HIP}$(\ovii)$\approx 3.4 \times 10^{16}$ cm$^{-2}$
and N$_{HIP}$(\oviii)$\approx 1.2 \times 10^{18}$ cm$^{-2}$). This
is noteworthy, since traditionally, the \ovii \ absorption has
been associated with that of \oviii \ and warm absorbers (e.g.
George et al. 1998 and references therein). \citet{ota96} have
previously suggested two different sources of \ovii \ and \oviii \
(for MCG-6-30-15), but their physical interpretation is completely
different than the one presented here, with the two absorption
components at completely different regions. In our model, \ovii \
is consistent with a cooler medium producing significant amounts
of O {\sc vi} and other lower ionization species. Without the
constraints introduced by the UTA this distinction could not be
realized. We also note that the O column densities produce
significant K-edge absorption ($\tau_{\lambda}=0.33$ for \ovii \
and $\tau_{\lambda}=0.13$ for \oviii).

Our model predicts more than 265 absorption lines with EWs larger
than 1 m\AA. As can be appreciated in Figure \ref{fig1}, most of
them fit the data in a quite satisfactory way. Table \ref{abl} and
Figure \ref{ew} also show a good agreement between the measured
and predicted EWs for unblended lines. The highly ionized phase
produces nearly all the features between 15.3 \AA \ and 8.5 \AA \
(Fig. \ref{fig2a}), where the spectrum becomes dominated by the Fe
L-shell lines, plus the contribution of high ionization lines
arising from other ions, such as O, N, and Mg. The predicted form
of the UTA by the cooler phase also shows a good agreement with
the data. We notice, however, that individual lines at around
$16.15$ \AA \ and $16.35$ \AA \ are not predicted by the model.
These features are part of the UTA (as can be observed in Figure 4
of Behar, Sako, \& Kahn; 2001), but the abbreviated data cannot
reproduce them. Our model does not yet include data for the O {\sc
vi} inner shell transitions. However, we notice that these
features can be observed in the spectrum, particularly the most
prominent one at 22.04 \AA. Unfortunately, the signal to noise
ratio at these wavelengths is too low to confirm their detection,
even at the 1 $\sigma$ level (see also Fig. \ref{fab}). This is
also the case of the feature at 22.33 \AA, which could arise from
O {\sc v}. However, the presence of these features in the spectrum
of NGC 3783 has been established through observations with the
XMM-Newton (Blustin et al. 2000). This is an important point,
since the LIP in our model predicts that a substantial fraction of
the O will exist in these lower charge states (N(O {\sc
vi})$\approx 7 \times 10^{17}$ cm$^{-2}$ and N(O {\sc v})$\approx
2 \times 10^{17}$ cm$^{-2}$).

There is another region where a discrepancy deserves comment.
Between 6.75 and 7.0 \AA, two lines clearly show larger EWs than
the predictions of the model. These intriguing lines may be
identified provisionally as Si {\sc x}($\lambda 6.850$) and Si
{\sc xi}($\lambda 6.775$). To reproduce them, either the high
ionization phase would have to be much cooler ($T<2\times10^5$ K)
with a much smaller value of the ionization parameter ($\log
U<0.5$), or the low ionization phase would have to be much hotter
($T>9\times10^4$ K) with a much larger value of the ionization
parameter ($\log U>-0.2$). As will be discussed in \S \ref{uta},
such values are not consistent with other features observed in the
spectra. The presence of a third absorber with intermediate U is
also unlikely  (see Fig. \ref{si}d). A possible explanation for
the difference is that our atomic database is incomplete,
containing data for only the most intense lines of Si (lines with
oscillator strengths larger than 0.1), so that the observed
features are in fact blends of several Si {\sc x} and Si {\sc xi}
lines. Alternatively, the atomic data used to calculate the
ionization balance for iron and silicon might be inconsistent.

Our model uses the abbreviated data set to fit the 15-18 \AA \
UTA. Since this is the main feature of the cooler phase, it is
important to know the effects that this limited data could have on
our results. We believe the approximation is sufficiently reliable
in this case for two reasons. First, we are using turbulent
velocities $> 100$ km s$^{-1}$ and for all ions of iron the column
density is lower than $10^{17}$ cm$^{-2}$ (which according to
Behar, Sako, \& Kahn (2001), are the limits within which the
approximation is valid, see \S \ref{code}). Second, our results
are in qualitative agreement with those by Blustin et al. (2000)
who modelled the UTA with the complete data set, and find a cooler
phase consistent with ours (see \S \ref{comp}). Therefore, it is
unlikely that an overestimate of the absorption is present, but
this still has to be confirmed. A quantitative estimation of the
uncertainties introduced by the abbreviated data will be presented
in a forthcoming paper.

\subsection{Ionization constraints on two, three, and multi-component models \label{uta}}

\begin{figure}[!t]
\figurenum{12} \plotone{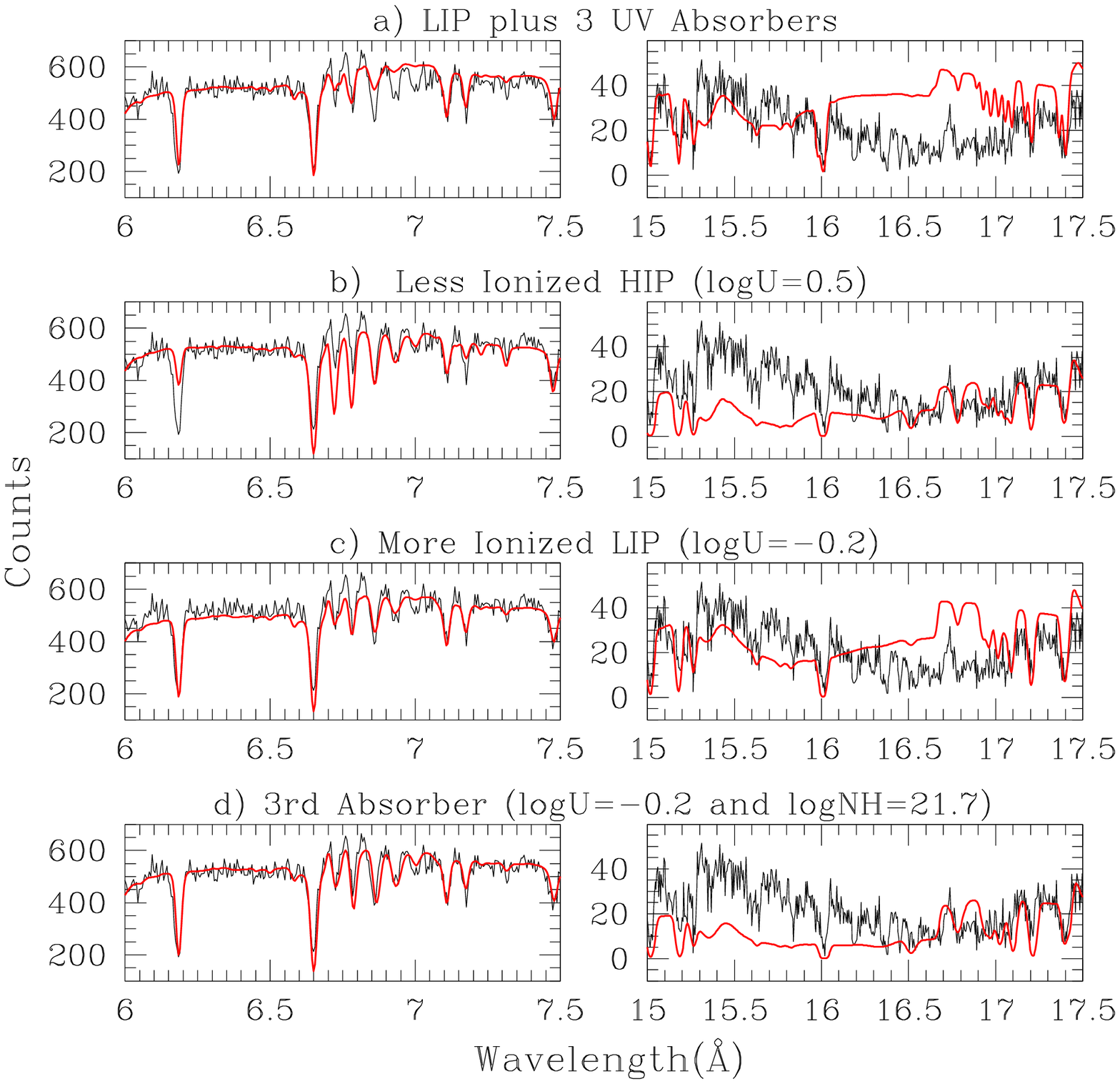} \caption[f13.eps]{\footnotesize
Panel (a): Low ionization component decomposed in three
subcomponents matching all the parameters predicted for the UV
absorbers (Kraemer et al. 2001). This fit shows several
discrepancies with the data. Middle panels: Different values for
the ionization parameters for our two phase absorber showing
inconsistencies with the observed UTA.  Panel (b): $\log$U=0.5 for
the high ionization phase. Note that the absorption from Si {\sc
xiv} is underestimated. Panel (c): In this case, the ionization
parameter of the low ionization component has been set to
$\log$U=-0.2. Panel (d): A third absorber has been included in our
model with $\log$U=-0.2 and $\log$N {\sc h}=21.7. The inclusion of
a new component could account for the missing Si lines, but would
overpredict the absorption by the UTA, therefore a flow with
continuous variation of U is ruled out.$^9$ \label{si}}
\end{figure}

\begin{figure}[!t]
\figurenum{13} \plotone{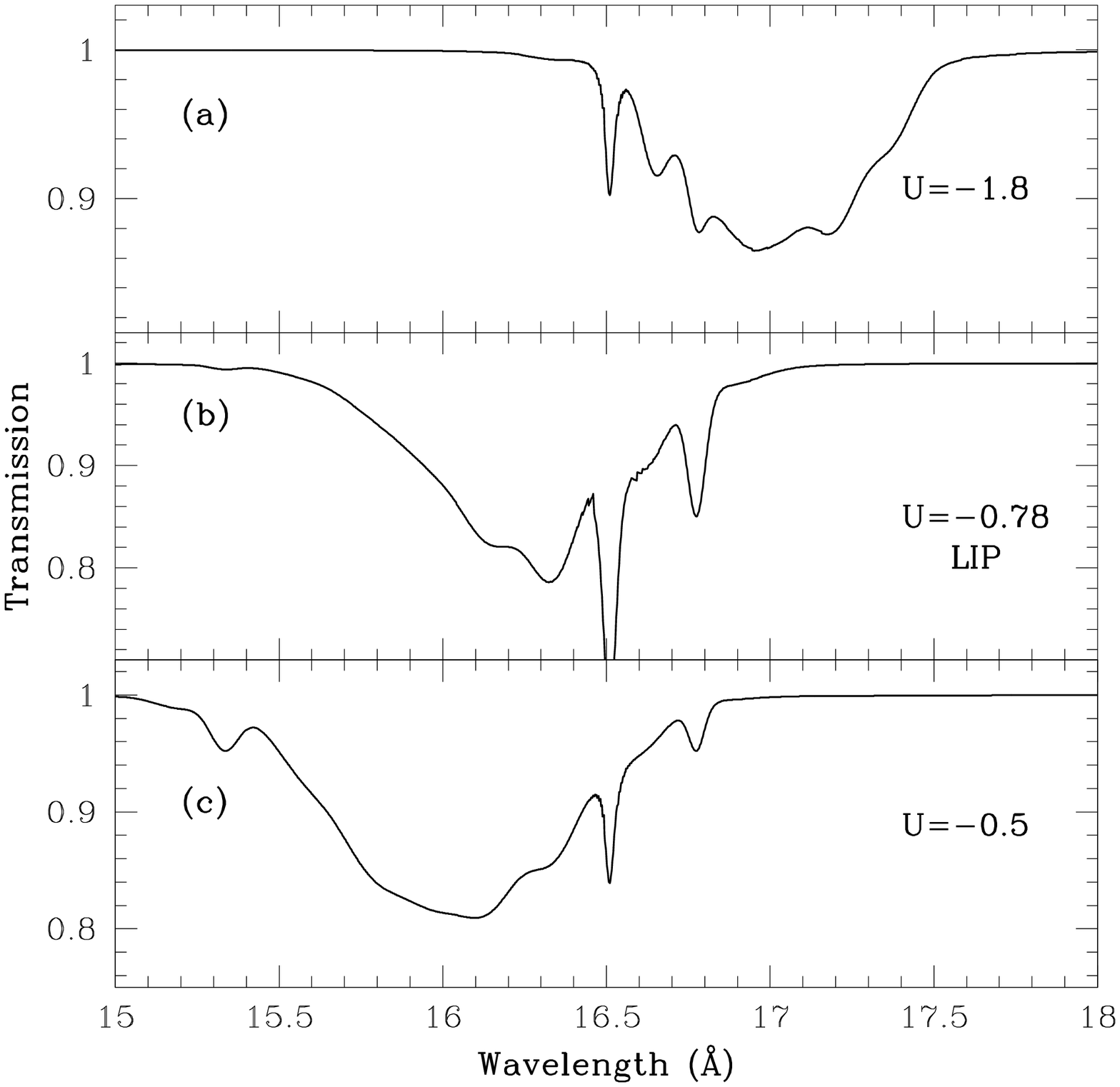} \caption[f14.eps]{ UTA model
predictions for different values of the ionization parameter. The
shape of the UTA is very well defined by the degree of ionization,
which results in important constraints to the physical conditions
of the ionized gas (Behar, Sako \& Kahn, 2001). \label{futa} }
\end{figure}

Since it is a wide, well resolved feature, the Fe M-shell UTA sets
tight restrictions on the ionization level of the gas (see Fig.
\ref{si}). As can be observed in Figure \ref{futa}, different
values of the ionization parameter lead to unambiguously different
shapes for the UTA. For instance, $\log U=-1.8$ will produce a
feature from Fe {\sc iii-vii} between 16.5 and 17.5 \AA, with the
peak of absorption around 17 \AA \ (Fig. \ref{futa}a). Lower
values of $\log U$ will give rise to a narrower UTA, centered at
wavelengths larger than 17 \AA. On the other hand, larger values
for the ionization parameter will shift the peak of absorption to
shorter wavelengths, and will produce a wider shape. A logarithm
of the ionization parameter equal to -0.78, like the one predicted
from the LIP in our model, produces absorption between 15.7 and
16.9 \AA, with an absorption peak at around 16.4 \AA \ (Fig.
\ref{futa}b). A further increase in the ionization parameter value
will further shift the peak of absorption toward smaller
wavelengths and will widen even more the shape of the UTA (Fig.
\ref{futa}c).

The UTA can thus be used to estimate a lower limit to the
ionization parameter of additional ionization states in the
absorber. The spectrum of NGC 3783 shows no trace of the inner
shell Fe {\sc xiv-xvi} M-shell transitions, and hence most of the
iron in the HIP has to be at least 15 times ionized. The fractions
of Fe {\sc xv} and Fe {\sc xvi} for this component are less than
$\sim0.003$ and $\sim0.001$ respectively.

The ionization parameter of the HIP is not consistent with a
smaller value capable of reproducing the Si absorption features
between 6.75 and 7.0 \AA, since it would produce a significant
fraction of Fe {\sc xii-xv}, and a UTA similar to the one
presented in the lower panel of Figure \ref{futa}, a feature not
observed in the spectrum (see section \S \ref{uta}). Furthermore,
a larger value would underpredict the absorption by Si {\sc xiv}
and other high ionization species. The resulting inconsistencies
can be observed in Figure \ref{si}b. The low ionization phase is
similarly well constrained. Figure \ref{si}c shows the effects of a
larger ionization parameter. In this case, the absorption by the
UTA at wavelengths larger than 16.5 \AA \ is underpredicted by
$>50$ \%.

The inclusion of another absorber with intermediate ionization
parameter and comparable H column density is also inconsistent
with the data. A third component, capable of reproducing the Si
lines, would overpredict the UTA absorption between 15.3 and 16
\AA \ by $>75$\%. Such an absorber would need an ionization
parameter $\log$U $\approx -0.2$ and a column density $\log$N$_H
\approx 21.7$. Figure \ref{si}d presents the effects of including
a third absorber. These constraints strongly suggest that no
intermediate values of U with high column density  are present.
However, the presence of the Si  {\sc x} and {\sc xi} absorption
features not accounted for in our model does not allow us to
effectively rule out this possibility. Nevertheless, the use of
better data for both the UTA and the Si inner shell transitions
may account for the discrepancies. We cannot rule out the presence
of additional low column density components or a bumpy continuous
distribution with a small contribution to the absorption by low
column density gas. However, such putative gas would have to have
small column densities ($\log$N$_H < 21$ [cm$^{-2}$]).
Hence, from the above analysis it is clear that only two
dominant absorbing components are required by the data in the MEG
wavelength range.

As a final remark, we would like to stress that the discussion
presented in this section is not dependent on the fact that we are
using the abbreviated data to model the UTA. This approximation
affects only sharp narrow edges. Figure 4 of Behar, Sako \& Kahn
(2001) presents three UTA models produced by media with different
ionization parameters carried out with the full set of Fe M-shell
transitions. A close inspection of their figure leads to the same
conclusions presented here.

The physical conditions derived for the ionized absorber in NGC
3783 are thus well constrained.

\begin{figure}[!t]
\figurenum{14} \plotone{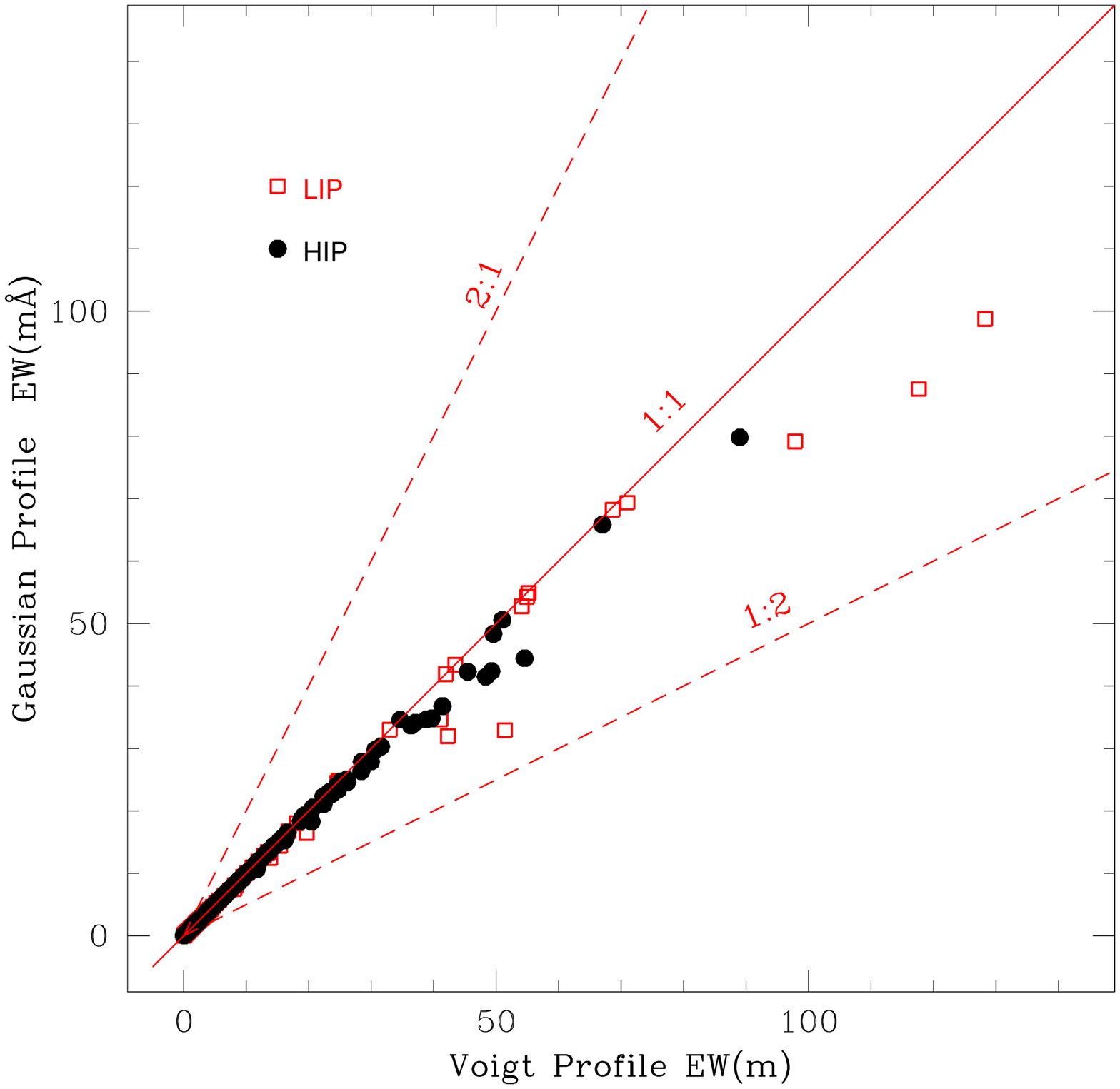} \caption[f15.eps]{Model predicted
EWs calculated using a Voigt profile vs. calculated using a
gaussian profile. The absorption lines from the HIP and the LIP
are presented separately. Deviation from the 1:1 identity line
indicates strong saturation.$^9$ \label{voigt}}
\end{figure}

\subsection{Line Saturation}

Our model predicts saturation of several absorption lines (see for
instance Fig. \ref{ble}). This saturation is not evident to the
eye due either to heavy line blending or to emission and
absorption line filling (see \S \ref{aeb}). Figure \ref{voigt}
shows a comparison of the EWs predicted from our model calculated
with a gaussian and a Voigt profile. Fewer than 20 lines show
saturation. However, as shown in the figure, the most intense
absorption lines from the LIP are saturated. From Figure
\ref{fig2b}, it is evident that the LIP contributes with fewer
absorption lines (but is responsible for most of the
photoelectric absorption) than the HIP. So, although present in only a
few absorption lines (the most intense ones) saturation 
becomes a significant effect.

\subsection{Absorption and Emission Blending \label{aeb}}

As can be observed in Figure \ref{fig1}, the model predictions
are in fairly good agreement with the data for the most intense
\ovii \ and \oviii \ absorption and emission lines. However, an
inspection of these lines from Table \ref{abl} shows that the EWs
derived by the model have systematically larger values than those
measured from the observations. We believe this effect is a result
of the filling of the absorption lines by the emission features.
In Figure \ref{ble}, we illustrate an example of the blending of
absorption and emission lines. Therefore, one really needs to take
into account absorption and emission together to correctly infer
the physical conditions of both the absorber and the emitter. In
our case, the emission was not inferred in a self-consistent way,
although, it was constrained by the absorption. Therefore our
emission fits (Table \ref{eml}) can be used as a rough estimate of
the emission line properties of NGC 3783. An inevitable conclusion
is that without modelling of absorption and emission line
blending, measurements of the observed EWs from the data will
underestimate the actual values. This effect can be observed in
Figure \ref{abem}a, where we present a comparison of the emission
line EWs with and without the contribution from absorption. This
figure also shows that a factor of 2 for the filling of the
absorption lines by the emission lines is present in almost all
the resonant transitions (the few forbidden transitions show no
effect, as expected).

\begin{figure}[!t]
\figurenum{15} \plotone{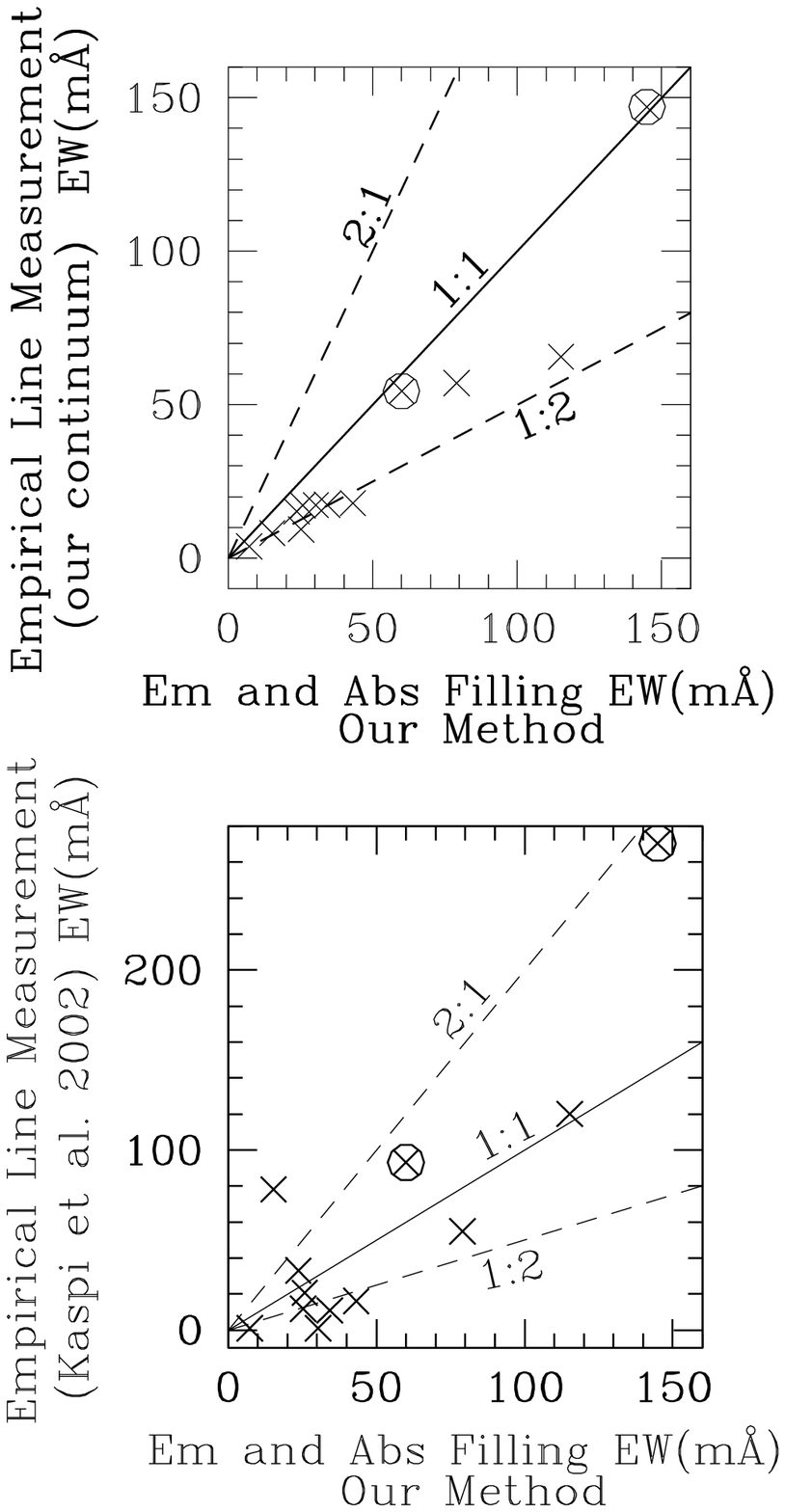} \caption[f16.eps]{\footnotesize
EW comparison for emission lines. Crosses within a circle stand
for \ovii \ forbidden and intercombination lines, where no
absorption and emission filling is expected. Upper panel: EWs
comparison between measurement taking and without taking into
account blending and absorption/emission filling model
predictions. Lower panel: EWs from Kaspi et al. (2002) vs. EWs
from this paper. In this case, since the emission/absorption
blending is not taken into account, and only a local continuum is
fitted, inconsistent results are obtained. The solid line has a
slope equal to 1. The dashed lines represent deviation factors of
1/2 and 2.\label{abem}} \end{figure}

Additional complexity is introduced to the absorption/emission
blending problem by the continuum level: if this level is not
modelled globally, inconsistent results can be obtained. In Figure
\ref{abem}b, we compare the emission properties from this paper
with those deduced by Kaspi et al. (2002), where no filling
correction was applied, and only a local continuum was fitted. For
truly unblended lines (such as \ovii \ forbidden and
intercombination lines), the values reported by Kaspi et al. are
larger than ours, indicating a lower continuum level. By contrast,
some emission lines blended with an absorption component have
lower EWs in their study (in some cases by a factor 2), a
signature of absorption filling. Even more, the fitting of a local
continuum can yield highly misleading results. For instance, Kaspi
et al. estimated the EW of the resonant Ne {\sc
ix}($\lambda$13.699) line as 78.1$\pm12.6$ m\AA, while we computed
a value of only 15.3$^{+14.6}_{-9.8}$ m\AA. Although this line is
filled by absorption, Kaspi et al. found a much larger EW than we
do, because they assumed that most of the spectral features near
this wavelength were produced by emission (i.e. they fitted a low
continuum level). As can be seen in Figure \ref{fig1} the
continuum level at this point from our global fit is rather high,
and most of the structures there are produced by absorption, with
only a small contribution from this emission line. This is a
common problem also faced in analyses of the Lyman $\alpha$ forest
(e.g. Bechtold et al. 2002).

\begin{deluxetable}{lccccccc}
\tablecolumns{8} \tablewidth{0pc} \tablecaption{Comparison with
previous models \label{pm}}

\tablehead{\colhead{} & \colhead{$\Gamma$} & \colhead{Norm.$^a$} &
\colhead{log(U$_{OX}^{LIP}$)} & \colhead{log(U$_{OX}^{HIP}$)} &
\colhead{log(N$_H^{LIP}$)} & \colhead{log(N$_H^{HIP}$)}
&\colhead{z(x)/z$_o$} } \startdata
This paper & 1.53$^b$ & .011 & -2.77 & -1.23 & 21.6 & 22.2& 1 \\
Kaspi et al.& 1.77 & 0.015 & -1.75 & 0.75 & 22.2 & 22.2&1 \\
Blustin et al&1.53& .018 &0.3$^c$&2.4$^c$&20.7& 22.45 &10(Fe) \\
de Rosa et al.& 1.83$^d$ & 0.019 &\nodata & 0.4 &\nodata & 22.3&1 \\

\enddata
\tablenotetext{a}{In photons keV$^{-1}$ cm$^{-2}$ s$^{-1}$ at 1
keV.} \tablenotetext{b}{The continuum includes the contribution
from a black body component with kT=0.1 keV, and norm=0.0002
$L_{39}/D_{10}^2$, where $L_{39}$ is the source luminosity in
units of $10^{39}$ erg s$^{-1}$ and $D_{10}$ is the distance to
the source in units of 10 kpc.}

\tablenotetext{c}{The value reported is $log\xi$, where
$\xi=L/r^2n$, $L$ the luminosity of the source, $r$ the distance
from the source to the cloud, and $n$ the cloud density. A
conversion to U$_{OX}$ was not possible since the authors did not
provide the SED.}

\tablenotetext{d}{The continuum includes the contribution from a
black body component with kT=0.21 keV, and norm=0.00045
$L_{39}/D_{10}^2$.}
\end{deluxetable}

\subsection{Spectral Energy Distribution \label{sedep}}

\citet{ma94} have shown the importance of including the particular
SED of an AGN, rather than a typical SED. In the latter case,
incompatible physical conditions may arise in the description of a
photoionized cloud. As discussed before (\S \ref{con1}), the SED
we used to calculate the ionization balance of the gas is only an
approximation to the actual (unknown) SED. This approximation is
the best that can be obtained currently for the spectral shape of
NGC 3783. This point is important, since the values obtained for
the ionization parameters are dependent on the continuum shape,
and moreover, so is the {\it difference} between the ionization
parameter of the two phases.

Our chosen SED (Fig. \ref{fsed}) produces rapid changes of the
ionization balance with small changes of the ionization parameter,
and therefore the ratio of U between the two phases is small when
compared to other SEDs (for instance the SED used by Kaspi et al.
2001). To illustrate this point, we have recalculated the
ionization parameters using the Mathews \& Ferland (1987)
continuum (shown in Fig \ref{fsed}}). The values obtained for this
SED for the low and high ionization component were $\log U=-0.05$
and $\log U=1.81$. These values are larger than those predicted
with our better chosen SED, and yield a factor $\approx$ 72
between the ionization parameter of the phases, twice as large as
we obtain. Therefore, any comparison between models should account
for the effects of the different SEDs used. In our case, a factor
of only 35 in the ionization parameters produces dramatically
different ionization degrees, with a factor $\approx 37$ between
the temperature of the two phases. However, using the Mathews \&
Ferland continuum gives T(HIP)=8.1$\times10^5$ K and
T(LIP)=2.1$\times10^4$ K, and so a difference factor in
temperature of $\approx 40$, all of which are close to the factor
found with our SED.

\subsection{Previous Models \label{comp}}

Four different studies have been published in the past for NGC
3783: (1) Kaspi et al. (2001) modelled a much lower quality
spectrum, the first HETG-{\it Chandra} observation of NGC 3783;
(2) Blustin et al. (2002) modelled also lower quality data, a 40
ksec exposure taken with the XMM-Newton; (3) de Rosa et al. (2002)
modelled low spectral resolution BeppoSAX observations; and (4)
Kaspi et al. (2002) analyzed the same 900 ksec HETG spectrum with
an empirical approach. In the following, we review these analyses
and contrast them with our results. Table \ref{pm} summarizes the
results found with the different models (including the one
presented in this paper).

\subsubsection{The model by Kaspi et al. 2001}

These authors established for the first time that two absorption
systems were required to fit the wide range of ionization species
found in the data. They estimated the continuum SED, including (1)
spectral curvature resulting from bound-free absorptions due to a
single component ionized absorber and (2) ``Compton reflected''
continuum. Using this SED, they then modelled the absorption
assuming the same turbulent and outflow velocities for the two
components (300 km s$^{-1}$ and 610 km s$^{-1}$ respectively).
They found a photon index $\Gamma=1.77$ and a normalization value
A(1keV)=0.0151 photons keV$^{-1}$ cm$^{-2}$ s$^{-1}$. While for
their low ionization component they found $\log U_{OXYGEN} =
-1.745$, for the high ionization one they obtained
$\log{U_{OXYGEN}} = -0.745$. In both components $\log$
N$_H$[cm$^{-2}$] was estimated to be 22.2.

A comparison between this model and ours (see Table \ref{pm})
indicates a difference of 0.24 between the photon indexes deduced
($\Gamma=1.53$ in our case). Compton scattering (not included in
our model) is unlikely to contribute a lot to this difference;
Kaspi et al. (2002) estimated it to be $\lesssim 15$\%. Rather,
the source of the discrepancy most likely arises from Kaspi et al.
(2001) basing their column density measurements only on the
photoelectric edges. The 900 ksec spectrum clearly shows that the
two deepest edges, \ovii \ and \oviii, are masked by Fe
transitions: \ovii \ is hidden in the Fe M-shell UTA (not modelled
by Kaspi et al.), and \oviii \ is hidden in the Fe L-shell lines.
Since the contribution of these lines was not considered, the edge
depths were overpredicted (see Fig. \ref{smita}), which, in turn,
yielded an overprediction of the photon index. This effect has
repercussions on any measurements based solely on low resolution
CCD spectra. In the case of NGC 3783, based on ASCA data,
\citet{rey97} and \citet{geo98} found $\tau_{\lambda}=1.2$ for
\ovii \ and $\tau_{\lambda}=1.4$ for \oviii, and thus a factor at
least 4 times larger than the edges predicted by our model. Figure
\ref{smita} shows a comparison between models including only
photoelectric absorption and models including both photoelectric
and resonant absorption. Therefore the depths and the column
densities calculated through low resolution spectra, should be
considered only upper limits to the actual values. In objects like
NGC 3783, where a deep UTA is present, the overestimation is
likely to be significant. We do not yet know the importance of
this effect in other objects.

To make a comparison of the ionization parameters, we calculated
$\log U_{OXYGEN}$ (the ionization parameter in the range 0.538 and
10 keV, George et al. 2000) for the continuum we used. The values
we obtained were -2.77 for the LIP and -1.23 for the HIP, factors
of 10 and 3 lower than Kaspi et al. (2001) values. While our HIP
has an ionization degree between the high and the low ionized
components of Kaspi et al., our LIP is much less ionized than
either. In both models, the low ionization phase is responsible
for the \ovii \ absorption, but the predicted physical conditions
of the gas are completely different. The difference in ionization
degree is explained by the omission of the UTA in their model. As
discussed in \S \ref{uta}, the ionization parameter of their low
ionization component is too high to reproduce this feature.

\subsubsection{The model by Blustin et al. 2002}

As in the previous case, Blustin et al. (2002) needed two absorption
components to
fit the XMM-RGS data. Blustin, et al. modelled the continuum with
a simple power law with $\Gamma=1.54$ and a normalization value
A(1keV)=0.0175. The normalization is larger than what we find,
because they have not included a blackbody component in their
model. The photon index, however, is consistent with the one we
found (see Table \ref{pm}). Their strategy for modelling the
absorption was to measure first the contribution of each ion with
the configuration {\it slab} of the code SPEX, and then to model
the data for a two phase ionized absorber using the configuration
{\it xabs}. The authors estimated the outflow velocity as 800 km
s$^{-1}$, and fixed a velocity width (rms) of 300 km s$^{-1}$.
This width refers to the gaussian sigma, and corresponds to a
Doppler velocity (turbulent plus thermal, see \S \ref{code})
$\approx 425$ km s$^{-1}$. The values predicted by their model
were $\log \xi=2.4$ erg cm s$^{-1}$ and $N_H = 2.8 \times 10^{22}$
cm$^{-2}$ for their high ionization phase, and $\log \xi=0.3$ erg
cm s$^{-1}$ and $N_H = 5.4 \times 10^{20}$ cm$^{-2}$ for the low
ionization one (where $\xi$, defined as $L/nr^2$, is the
ionization parameter). A direct comparison between ionization
parameters is not possible since the SEDs used are different (see
\S \ref{sedep}). However, our results are not qualitatively in
contradiction with theirs.

Nevertheless, there are several quantitative differences that lead
to different interpretations of the data: (1) While we were able
to fit the data assuming solar abundances, their low ionization
phase requires 10 times the solar abundance of iron to reproduce
the observed features. This value seems to be rather high. (2) The
H column density of their cooler phase is $\approx 8$ times lower
than ours. (3)  By contrast with our model, in theirs  the hotter
component has a significant contribution to the absorption
produced by  \ovii, due to the high column density of this
component (and the low column density of their cooler component).

These quantitative differences can be understood from the
different continuum fits applied to the data, i.e. due to the
inclusion of a soft thermal component in our model. The presence
of a soft thermal component requires deeper
absorption at long wavelengths to match the data. Therefore, a cooler
phase capable
of accounting for this absorption needs to have a high H column density.
However, this would also result in strong saturation in the C, N,
and O lines. Since, as stated by Blustin et al.,  the
abundance of Fe is effectively measured against the abundance of
these ions, omitting the soft excess in the
presence of a deep UTA would require an overabundance of Fe
relative to other elements. Therefore, Blustin et al. concluded
that iron was around one order of magnitude more abundant than
oxygen (relative to solar values) and, accordingly, that the H
column density was around one order of magnitude smaller in the
low ionization phase because, at the time, they did not find evidence
for a soft excess. However, they emphasize that if saturation
is present (along with the presence of a soft excess), the iron
abundance could be much smaller. This is in agreement with our
results, which require Fe/O to be within a factor 2 of solar and
strong saturation. The inclusion of a thermal component,
therefore, allowed us to fit the data with solar abundances.

The existence of a soft excess in the spectrum of NGC 3783 has been
further evidenced by the X-ray variability properties seen with
new XMM data. The overall 0.2-10 keV spectrum gets softer as the
source gets brighter, consistent with a variable soft excess, as
seen in other Seyfert galaxies (A. Blustin, private
communication).

\subsubsection{The Model by de Rosa et al. 2002}

BeppoSAX observed NGC 3783 5 days in June of 1998. De Rosa et al.
2002 reported this observation and modelled the data. Only modest
flux variations ($\sim$ 20\%) were detected. They found a photon
index $\Gamma =1.83$ and a normalization value A(1keV)=0.019
photons keV$^{-1}$ cm$^{-2}$ s$^{-1}$. A thermal component with
$k$T= 0.21 keV and normalization 4.5$\times 10^{-4}$
$L_{39}~D_{10}^{2}$ (where $L_{39}$ is the source luminosity in
units of $10^{39}$ erg s$^{-1}$ and $D_{10}$ is the distance to
the source in units of 10 kpc) was necessary to fit the spectrum.
They also found evidence for the ionized absorber. The results of
their fitting are presented in Table \ref{pm}. As in the case of
Kaspi et al. (2001), the differences between fittings can be
explained by the fact de Rosa et al. used only the absorption
edges (due to the low resolution of their data), and not the
lines, in their analysis.

\begin{deluxetable}{lcc}
\tablecolumns{3} \tablewidth{0pc} \tablecaption{Comparison with
Kaspi et al. (2002) \label{kas}}

\tablehead{ \colhead{$\lambda$ (\AA)}& \multicolumn{2}{c}{Ion Name
and Rest Frame $\lambda$ (\AA) identification} \\
\cline{2-3}  \colhead{Observed} & \colhead{Kaspi et al.} &
\colhead{This paper} } \startdata

10.126 &  Fe{\sc xvii}(10.112) & Fe{\sc xvii}(10.112), Fe{\sc xix}(10.119), Fe{\sc xvii}(10.120) \\
10.524 &  Fe{\sc xvii}(10.504) & Fe{\sc xvii}(10.504), Fe{\sc xviii}(10.537) \\
12.436 &  No identification & Ni{\sc xix}(12.435) \\
12.560 &  Fe{\sc xx}(12.576) & Fe{\sc xx}(12.526) \\
12.592 &  Fe{\sc xx}(12.588) & Fe{\sc xx}(12.576,12.588), Fe{\sc xix}(12.538) \\
13.612 &  \nodata & Fe{\sc xix}(13.645,13.643) \\
13.822 &  Fe{\sc xix}(13.795) & Ne{\sc vii}(13.814),Fe{\sc xix}(13.795) \\
14.269 &  Fe{\sc xviii}(14.256) & Ne{\sc v}(14.239),Fe{\sc xviii}(14.256) \\
15.584 &  \nodata & Fe{\sc xviii}(15.625) \\
21.466 &  O{\sc vii}(21.602) zero redshift & Ca{\sc xvi}(21.450) \\
23.783 &  \nodata & Ne{\sc vi}(23.771) \\
24.124 &  \nodata & Ca{\sc xiv}(24.114) \\
\enddata

\end{deluxetable}

\subsubsection{Analysis by Kaspi et al. 2002}

Kaspi et al. (2002) used a different analysis strategy and did not
fit the data with a theoretical model. They empirically identified
the lines and measured the EWs directly from the spectrum, using
apparent line free zones to estimate the continuum level, but only
locally.  The appeal of this approach is that it produces
directly measured quantities before getting into detailed
modelling. The disadvantages are: a conflating of blends into
single ``features'', the ignoring of the mutual cancelling of
emission and absorption, and an inability to derive a global
picture of the physical conditions.

There are some discrepancies between their line identifications
and the predictions from our model. The most important ones are
listed in Table \ref{kas}. The differences can be attributed
mostly to the inclusion in our calculations of Ni and Ca lines, as
well as more than 1200 Fe transitions arising from the ground
state. According to Kaspi et al. (2001) their models include
several hundred. The authors tentatively identified a 14$\pm16$
m\AA \ absorption line around 21.45~\AA \ (in the rest frame
system, Fig \ref{fig1}) as a local component of \ovii ($\lambda
21.602$). This feature would be produced by intergalactic
absorption at zero redshift, or by Galactic absorption (see
Nicastro et al. 2002). Our model indicates that this line is
produced by the high ionization phase, and corresponds to a Ca
{\sc xvi}($\lambda 21.45$) transition (the difference in
wavelength is due to the outflow velocity of the absorber).
Therefore, any putative contribution from a local medium cannot
account for an EW larger than 9.3 mA (at 1$\sigma$ level),
consistent with the results found for the local filament by
Nicastro et al. (2002, 2003). An EW comparison between Kaspi et
al. and our measurements indicate a good agreement for all
unblended absorption lines (14 cases) and for some blended
features (11 out of 17 cases). The differences found can be
ascribed to the different continuum levels. While we fitted the
continuum derived from the global spectrum using our model, they
measured the continuum locally and in an empirical way. We notice
that, as discussed in \S \ref{aeb}, several absorption-emission
blended lines present systematic differences between measurements
and our model predictions.

\section{Absorber Physics Discussion \label{disc2}}

\subsection{Two phases of the same medium? \label{2phase}}

The simple nature of our two component model, with
indistinguishable kinematics and very different ionization
parameters, suggests that the absorption observed in the NGC 3783
spectrum may arise from two phases of the same medium.

First, two separate components seem to be the best way of
describing the absorbing gas, since the discussion presented in \S
\ref{uta} and the plots in Fig. \ref{si}d strongly suggest that a
continuous medium, i.e, a medium with a smooth distribution of
temperatures from the HIP to the LIP (as proposed by Krolik \&
Kriss, 2001) is unlikely to produce the well defined UTA
absorption feature observed. Rather, such a medium would produce
less deep, continuous, absorption from $\approx15$ to $\approx 17$
\AA. A continuous medium also requires strong iron L and K edges
\citep{krol95}, features not detected in the spectrum.

\begin{figure}[!t]
\figurenum{16} \plotone{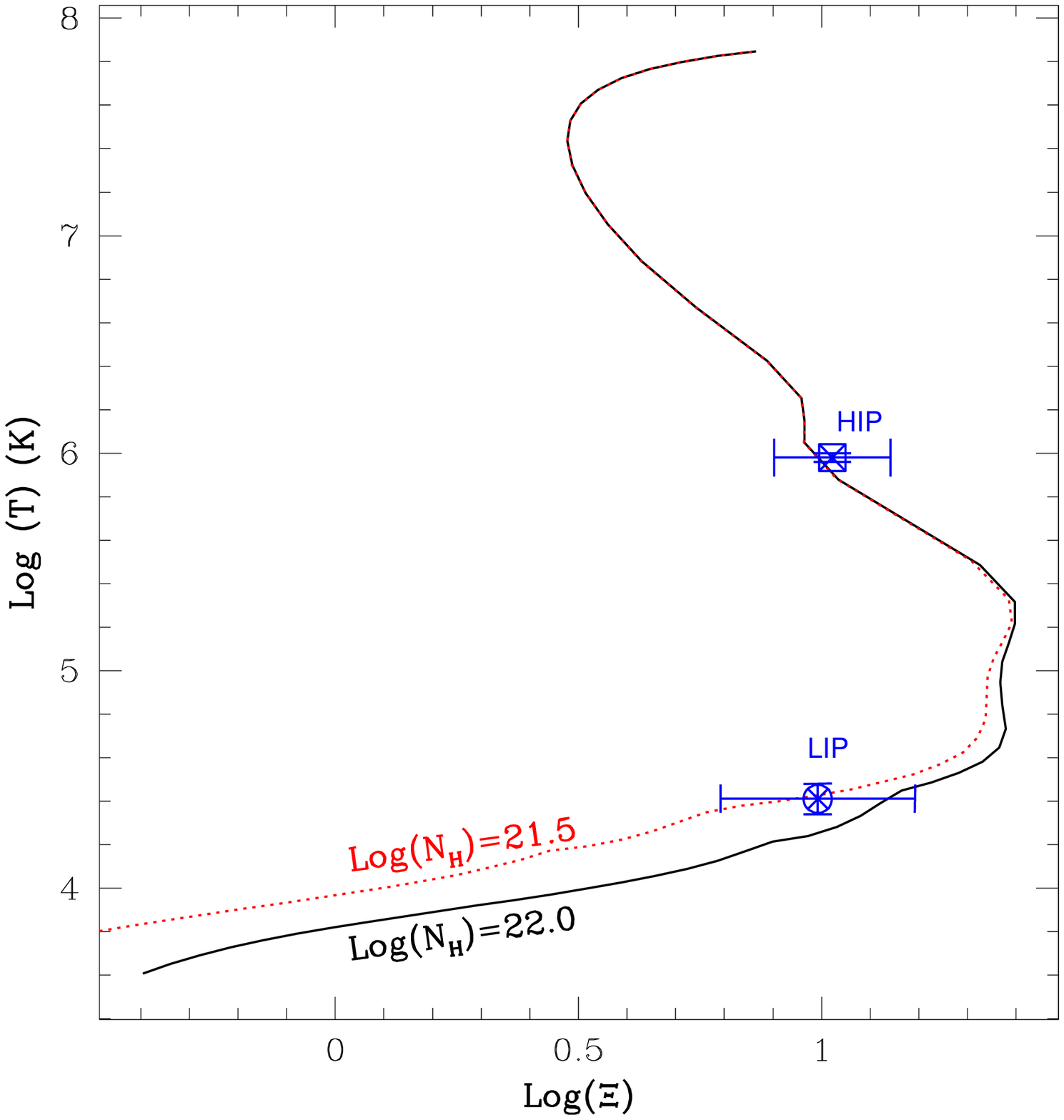} \caption[f17.eps]{Curve of
thermal stability for the SED used in this analysis, as described
in \S \ref{con1}. Pressure equilibrium between the HIP and the LIP
can be observed.$^9$ \label{xi}}
\end{figure}

The presence of a two phase medium (and not a continuous medium)
is further evidenced by the fact that the two components are in
pressure equilibrium: the gas pressure (P) of each phase is
proportional to the electron density (n$_e$) and the temperature
(T); therefore, with the adopted definition of the ionization
parameter, P$\propto$T/U. Under the assumption that the two
components lie at the same distance from the central source, as is
plausible given their identical kinematics, the two phases turn
out to have the same pressure (see Table \ref{2ab}). To further
illustrate this point, we show in Figure \ref{xi} the thermal
equilibrium curve (S curve) for the SED used in our analysis (see
\S \ref{con1}). To construct this curve we calculated the
ionization parameter as defined by Krolik, McKee \& Tarter (1981):
$\Xi=(L/4\pi r^2)/c(nkT_e)$ (where $L$ is the luminosity of the
source, $r$ the distance between the gas and the source, $c$ the
speed of light, $n$ the total particle density of the gas, $k$ the
Boltzmann constant, and $T_e$ the electron temperature). This
representation of the ionization parameter is simply the ratio between
the radiation pressure and the gas pressure. In Figure \ref{xi} we
also plot the $\Xi$ values we find for the HIP and the LIP (as listed in
Table \ref{2ab}). The pressure equilibrium is evident. Unless
imposed by real pressure equilibrium, this pressure equality would
have to be assumed to be a coincidence. We notice that the HIP
lies in the intermediate part of the curve, in a region where the
slope is negative. Regions of the curve with negative slope are
thermally unstable. This might be then inconsistent with the
equilibrium scenario just presented. Nevertheless, the shape of
the thermal equilibrium curve strongly depends on the SED, and
particularly on the unobservable region between 100 \AA \ and 912
\AA. Considering different shapes for the SED in this unknown gap
can give rise to thermal curves with an intermediate stable region
in the region occupied by the HIP (see for instance Reynolds and
Fabian 1995). Therefore, this
apparent instability might be only the result of our chosen SED, a
likely possibility, since according to the error bars in Figure
\ref{xi}, our HIP is consistent with the semi-stable intermediate
region of our S curve (nearly vertical region of the equilibrium
curve with a $\Xi$ value close to 1 and T $\approx 10^6$ K).
Alternatively, this instability could be reflecting the fact that,
besides the illumination from the central source, additional
cooling/heating processes (for instance shock-heating) might be
also working on the gas. In this latter case, the electron
temperature would be constant due to the additional source of
heating, and the solution found would be stable.

A two phase medium is consistent with the thickness ratio between
the phases: the radial thickness occupied by the absorbing gas is
roughly equal to the ratio between the H column density and the H
density (D=N$_H$/n(H)), and, from the definition of the ionization
parameter given in \S \ref{code}, D$\propto$N$_H$r$^2$U/Q(H).
Therefore the thickness ratio between the phases goes as
R$_{tck}$$\propto$ N$_H$(HIP)r(HIP)$^2$U(HIP)/
N$_H$(LIP)r(LIP)$^2$U(LIP). If we assume again that both phases
lie at the same distance from the central source, then the
thickness of the HIP is 140 times larger than that of the LIP.
This result is consistent with pressure confinement of the LIP by
the HIP.

A two phase warm absorber with no intermediate value of U was also
detected in IRAS 13349+2438 by Sako et al. (2001). In that case, the
UTA was also produced by Fe {\sc vii-xii}. Are the two phases in
IRAS 13349+2438 also in pressure balance? An intriguing
possibility is that a two phase medium forms because of the
thermal instabilities. In photoionization equilibrium the gas will
always be driven to the stable regions of the S curve which are set by
the SED and atomic physics. However,
not all the regions of the stable branches can be occupied by the
absorbing gas, as the two phases will tend to reach pressure balance
with each other.

This appealingly simple picture not only suggests the idea of a
two phase medium, but also points to a simple view of active
nuclei (e.g. the structure for quasars  suggested in Elvis 2000).
Recent results indicate that the physical properties  of the
optical `High Ionization Nuclear Emission Line Region' (HINER) are
consistent with those of the X-ray warm absorber \citep{erk97,
nic99b, net02}. Therefore, it is plausible that the high
ionization absorption and emission arise from the same gas. If the
gas lies in our line of sight, the absorption is observed, if the
gas lies out of our line of sight, only the coronal emission from
the HINER can be detected. \citet{por99} found that this picture
could only be supported if the electron densities are high
(n$_e\approx 10^{10}$ cm$^{-3}$) to avoid overproducing the
coronal lines, and concluded that the distance of the ionized
absorber from the incident radiation source is of the order of
that of the Broad Line Region (BLR). All this evidence also points
to models where the hotter gas confines the broad emission lines
gas (BELs gas, see, e.g. Turner et al. 1993; Elvis 2000).

Our two phase solution for the warm absorber may also be pointing
to this general scenario. The value of the electron density
derived from the \ovii \ triplet for NGC 3783 (see \S \ref{em}) is
consistent with the predictions by Porquet et al. (1999, see
above), and therefore this is consistent with a connection between
the absorbing and emitting gas. This density, as well as the
temperature for the HIP are close to those needed to explain
strong \ovii \ emission lines in other objects
\citep{geo95,nic99}. However, what is the location of the
absorber? We do not have any means of measuring the electron
density of the absorber or its distance to the central source at
this time. Nevertheless, \citet{nic99} and \citet{net02} have
shown through spectral variability studies in NGC 4051 and NGC
3516 respectively, that the location of the ionized absorbing gas
is similar or closer to the central source than that of the broad
line region.

A simple identification between the absorbing gas and an emission
component is difficult to obtain, but from the latter discussion a
natural candidate could be the Broad Emission Line region. This
has been already suggested by Kuraszkiewicz \& Green (2002), who
found a correlation between the N {\sc v}/C {\sc iv} ratio in
broad emission lines and that in narrow absorption lines, and is
further supported by the similarity of the temperature of the LIP
and the temperature predicted for the BEL (e.g. Kaastra et al.
1995, for NGC 5548). The BEL gas in AGN is clearly radially
stratified, as shown by reverberation mapping \citep{pet99}, with
a range of ionization parameters. According to the above models,
the BELs are produced by the contribution of {\em all} the clouds
surrounding the AGN, that can have different ionization conditions
(simply because of different distances to the central source, or
different electron densities), giving rise to the range of
ionization parameters. However, the absorption is produced only by
those few clouds intercepting our line of sight. Therefore, a
discrete value of the ionization parameter for the LIP can in
principle be reconciled with the LIP-BEL gas connection.
\citet{kora92} found that a log(U) range between -1.3 and -0.7
could characterize the UV BEL in NGC 3783. This range could be
consistent with the value obtained for the LIP. However, different
SEDs were used in the studies, and though they have similarities
(for instance, we included a change of slope in the UV, and they
included a big blue bump) it is not possible to make a clear
connection (see \S \ref{sedep}). Nevertheless, the similarity of U
(also found in other AGN, Elvis 2000), as well as the similarity
of the temperatures  are physically suggestive, and bear further
investigation.

This would point to models where the hotter emitting/ absorbing
gas confines the BELR/ LIP (see, e.g. Turner et al. 1993, Elvis
2000). Pressure confinement is currently not favored \citep{pet97}
among other reasons because of the rapid destruction of the BEL
clouds by drag forces, when they are moving through a stationary
medium. However, pressure confinement is possible if both phases
are comoving in a wind with a single outflow velocity, as is the
case of our two phases on NGC 3783. Other problems, such as the
need for a low Compton depth are solved if a funnel shape geometry
is present (Elvis 2000). This geometry is not in contradiction
with our results.

As a final remark, we note that the presence of the absorption
component detected in \ovii \ with an outflow velocity of 1365 km
s$^{-1}$ is not straightforwardly explained in light of the above
models. This component is well constrained, as discussed in
section \ref{hivel}, although contributing to only 4 features out
of the 265 with EW larger than 1 m\AA. However, the thickness
of the LIP is $\sim$9 times that of this component, and that of
the HIP is $\sim 1200$ times larger. Therefore, whatever the
origin of this absorption, it is not in contradiction with our
scenario in which two phases dominate. Nevertheless, what is
the true nature of this
puzzling component? The velocity matches that of the component
discovered in the UV, and in several other AGN multiple velocity
UV absorption systems have also been detected. Are these intervening
systems unrelated to the AGN? Do they
originate at the edge of an obscuring torus? The present data are
inadequate to say.

Our results, though strongly pointing to the general scenario
presented above, are based only on the analysis of one object. Was
this result only a coincidence? Or can all
ionized absorbers be represented with a simple two phase medium?
IRAS 13349+2438 is another example of an ionized absorber with two
different absorption components (Sako et al. 2001). Are these two
components always in pressure balance?  Further study is still
necessary to unravel the answer to these questions, but the simple
picture presented here is certainly intriguing.

\subsection{The Nature of the UV X-ray Absorption }

A warm absorber was first discovered by Halpern in 1984, but the
use of warm absorbers to understand the physical conditions in
matter close to the central black hole came with the realization
that the warm absorbers seen in X-rays also imprint signatures in
the UV as associated absorption lines. In a series of papers
Mathur and collaborators (1994 through 2001) showed that the high
ionization absorption lines seen in the UV, such as \ovi, \civ,
Ly$\alpha$ could come from the same single phase gas which
produced \ovii \ and \oviii \ edges in X-rays. The combination of
saturation free X-ray column densities with outflow velocities
from UV spectra led to the discovery of a component of nuclear
material not previously recognized: highly ionized, outflowing,
low density material situated close to or outside the broad
emission line region. The mass outflow rates implied are large, a
significant fraction of the mass accretion rate needed to power
the AGNs, and so must have important dynamical influences within
the active nuclei. The insight obtained through these studies led
to one of the most comprehensive models of the nuclear regions of
AGNs to date (Elvis 2000).

As such, the X-ray/UV absorbers have proved to be powerful tools
to understand the physics of AGN outflows. However, the connection
between UV and X-ray absorbers has remained a matter of
controversy for years, for several reasons:

(1) Intrinsic observational limitations make component
identifications difficult: even now, UV spectra have 20~times
better spectral resolution, so a simple, one to one, matching of
features cannot be performed. In addition, X-ray instrumentation
is sensitive to only a limited range of column densities (see Fig.
\ref{fab}), making weak components found in the UV undetectable,
even if present. For an ionization parameter U$\sim$ 1, FUSE can
detect 100 times smaller N$_H$ systems  (Kriss et al. 2000) than
even high S/N {\it Chandra} grating spectra (Fig. \ref{fab}). The
situation gets more complicated because only a limited range of
high ionization states can produce features in the UV spectra
(Fig. \ref{fab}), since the low ionization ion populations where
the UV transitions are produced become tiny. We now see that a
2-phase medium is required. The high ionization phase in our
model, as well as the high velocity system in NGC~4051 (Collinge
et al. 2001) represent good examples of this.

(2) As already stressed by Mathur et al. (1994, see also Kaspi et
al. 2001), the actual ionization parameters are strongly dependent
on the far UV shape of the SED. Although this does not affect the
absorption produced in the X-ray region (Steenbrugge et al 2003),
it has an important effect on the UV absorption and so on linking
the UV and X-ray absorbers. This is an important limitation since
this region cannot be observed due to Galactic absorption.

(3) The strong variability of the sources makes the determination
of the relative UV and X-ray continuum fluxes unreliable, unless
truly simultaneous observations are considered (Crenshaw et al.
2003).

(4) As discussed before, several studies on the UV have neglected
the effects of saturation, {\em under}estimating the ionic column
densities in the UV band. As demonstrated by Arav et al. (2003),
when an accurate treatment of saturation is present the column
densities found in the UV band become consistent with the X-ray
absorbing material. We find here that X-ray column densities have
instead {\em over}estimated the Oxygen edge depths and column
densities due to line blends.

(5) Finally, the presence of a hotter phase (not detectable in the
UV), as well as the presence of deep bound-free absorption edges
in the X-rays, introduces additional complexity. While the UV
absorbers are almost transparent one to the other, photoelectric
absorption in the X-ray attenuates significantly the continuum
each absorber receives. Hence, the order in which the absorbers
are illuminated by the source may affect the final results in a
given model. The ionization parameter is very sensitive to this
effect, making the comparison of UV and X-rays values unreliable.
Therefore, to really infer something about the X-ray UV absorption
link, simultaneous self-consistent modelling of each absorber
should be applied to simultaneously observed data. This is beyond
the scope of this paper (but we notice that simultaneous
observations with FUSE and HST-STIS do exist for the {\em Chandra}
data analyzed here).

Some of the difficulties mentioned above can be overcome with
careful analysis and modelling. But the complexity of UV and X-ray
spectra have led to many papers with highly complicated models
involving multiple physical components to describe AGN absorbers.
The general consensus from these studies, using very long
observations on the most powerful satellite borne telescopes,
seems to be that the absorption system(s) is/are extremely
complex. This renders warm absorber studies completely ineffective
as probes of circumnuclear environment of AGNs.

In this paper we have shown that the situation is much more
promising: the warm absorber in NGC 3783 can be simply described
by two gas zones, but no more, with identical kinematics within
the power of our measurements (\S \ref{750}), one having a low and
one having a high ionization parameter; these two zones are in
pressure equilibrium with each other and lie on the (T,$\Xi$)
equilibrium curve for the SED of NGC~3783, and so are two phases
of the same medium; the LIP phase also produces the UV absorption
lines (\S \ref{uv}).

Thus, the high resolution {\it Chandra} and XMM-Newton spectra can
resolve years of controversy surrounding the nature of X-ray/UV
absorbers, if analyzed and interpreted carefully. It appears clear
that (though multiple components may be present) the X-ray and UV
absorbers are different manifestations of the same outflow (e.g.
Steenbrugge et al. 2003), confirming the original Mathur et al. \
models, but refining them to a two phase medium. This is extremely
important, since the long baseline of ionization states from UV to
X-rays, and superior velocity information from higher resolution
UV spectra, can be used to model the absorbing outflow accurately.
We will exploit this tremendous potential in a forthcoming paper
presenting detailed model of the X-ray/UV absorber in NGC 3783,
and will extend the current analysis to more AGNs as high quality
X-ray spectra become available.

\section{Summary}

We have developed PHASE, a new code designed to model the X-ray
and UV absorption of ionized gas in the environment of AGN.  We
have used our model to study the {\it Chandra} 900 ksec spectrum
of NGC 3783. This X-ray spectrum is the best one available so far
from a Seyfert galaxy. We used a global fit approach, which is
necessitated by the blending of 75\% of the absorption features,
compromising line strengths measured empirically. The heavy
blending also results in a continuum level that cannot be
estimated locally, except in narrow bands. Our main results are as
follows:

(1) The intrinsic continuum of the source is well reproduced by a
power law ($\Gamma=1.53$) and a thermal component (kT=0.1 keV). We
attenuated this continuum by an equivalent hydrogen column density
of $1.013\times10^{21}$ cm$^{-2}$ to account for the Galactic
absorption.

(2) Our absorption model leads to a simple picture consisting of a
two phase wind. The model can reproduce more than 100 features
with only 6 parameters.  The high quality of the data, plus the
inclusion of new atomic data for several lines  made the modelling
results more reliable than previous studies. The two components of
the model are in pressure equilibrium, and are consistent with a
single outflow ($\approx 750$ km s$^{-1}$), a single turbulent
velocity ($ 300$ km s$^{-1}$), and solar abundances. The
ionization parameter of the high ionization phase (U=5.754) is
$\approx 35$ times larger than that of the low ionization phase
(U=0.166). The equivalent H column densities were estimated as
N$_H=1.6\times 10^{22}$ cm$^{-2}$  for the hotter phase and
N$_H=4.1\times 10^{21}$ cm$^{-2}$ for the cooler one. The
difference found for the ionization parameters of the phases is
strongly dependent on the continuum shape. With our chosen SED, a
factor of 35 yields a low ionization phase with a temperature 37
times lower than the high ionization one, and completely different
ionization degrees. The main features of the cooler phase are the
Fe M-shell UTA and the \ovii \ lines, while the features for the
hotter phase are the \oviii \ and the Fe L-shell lines. The \ovii,
previously identified with the \oviii \ and a hot phase, is
consistent with a cooler phase and the O {\sc vi}.

(3)Our model strongly disfavors a continuous range of ionization
parameters to describe the medium because it is not consistent
with the observed shape of the UTA.

(4) Thanks to the high resolution spectra, we were able to
determine that in the case of NGC 3783, the K edge of O {\sc vii}
is masked by the presence of the UTA and the K edge of O {\sc
viii} is masked by the presence of the Fe L-shell lines. Through
our model, we have shown that this produces an overestimation in
the measurement of the edge depths (by a factor $>$ 4), if the
contribution from the absorption lines is not taken into account.
An important conclusion derived from this is that only an upper
limit to the column densities can be obtained from CCD low
resolution spectra (e.g. ASCA). In objects where a deep UTA is
present, this effect could be particularly important.

(5) Our model predicts a Ca {\sc xvi} line at 21.45 \AA. This line
was formerly identified as a local \ovii ($\lambda 21.602$)
feature arising from an intergalactic cloud at zero redshift or
Galactic absorption. With the inclusion of the Ca line, if a local
feature is present, it cannot contribute more than 9.3 m\AA \ to
the total EW.

(6) Our model does not attempt to predict the emission lines in a
self-consistent way. Therefore, we fitted them with gaussian
profiles. We restricted the outflow velocities and FWHMs of the
lines arising from transitions of the same ion. The emission and
absorption lines are blended, making an independent estimation of
them unreliable. Our emission lines are constrained by the
absorption lines, making our predictions useful as rough estimates
of the actual emission line properties of NGC 3783. The emission
line EWs increase by a factor $\sim$2 when the absorption/emission
filling effect is included.

(7) We find an excellent agreement between X-ray data and a model
with kinematic components identical to the UV absorption lines. We
also find that the UV value of ionization parameter underpredicts
ionic column densities by as much as a factor of ten, illustrating
the danger of modelling the absorbers using UV data alone. A
detailed model of the UV and X-ray data together, however, is
beyond the scope of this paper.

The analysis presented in this paper offers new hope that warm
absorbers in AGN are relatively simple systems with physical
properties that are easily understood. If this turn out to be
generally true, then they will provide, particularly through their
variability, completely characterized physical conditions for AGN
winds. This in turn is likely to lead back to a deeper
understanding of the basic physics powering AGN and quasars.

\acknowledgements We thank the anonymous referee for constructive
comments which helped to improve the presentation of the paper.
This research has been partly supported by the
CONACyT grant 138012, NASA Contract NAS8-39073 (Chandra X-ray
Center) and Chandra General Observer Program TM3-4006A. Work at
LLNL was performed under the auspices of the U.S.\ Department of
Energy by the University of California Lawrence Livermore National
Laboratory under contract No.\ W-7405-Eng-48.

\end{document}